\documentclass[%
reprint,
superscriptaddress,
amsmath,amssymb,
aps,
prx,
]{revtex4-2}

\usepackage{graphicx}
\usepackage{dcolumn}
\usepackage{bm}
\usepackage[colorlinks]{hyperref}
\usepackage{physics}
\usepackage[squaren]{SIunits}
\usepackage{mathtools}
\usepackage{amsthm}
\usepackage{xcolor}
\usepackage{soul}
\usepackage{mathrsfs}
\usepackage{times}
\usepackage{comment}
\usepackage[normalem]{ulem}
\usepackage{algorithm2e}
\usepackage{titletoc}
\usepackage[colorlinks]{hyperref}
\usepackage{makecell}
\usepackage{tabularx}

\usepackage{tocloft}

\hypersetup{
	colorlinks=true,
	linkcolor=red,
	filecolor=blue,      
	urlcolor=blue,
	citecolor=blue
}
\newcommand{\nisqrc}{NISQRC}
\newcommand{\NS}{S}

\newcommand{\UI}{\mathbf{u}}
\newcommand{\YO}{\mathbf{y}}
\newcommand{\YS}{\mathbf{y}^{\star}}
\newcommand{\Es}[1]{\mathbb{E}_{\mathcal{X}}\!\!\left[#1\right]}

\newcommand{\Eq}[1]{Eq.\,(\ref{#1})}

\newcommand{\rhofpm}{\hat{\rho}^{\mathsf{M}}_{\mathrm{FP}}}

\begin{document}


\title[]{Overcoming the Coherence Time Barrier in Quantum Machine Learning on Temporal Data}

\makeatletter

\author{Fangjun Hu}
\thanks{These two authors contributed equally}
\affiliation{Department of Electrical and Computer Engineering, Princeton University, Princeton, NJ 08544, USA}

\author{Saeed A. Khan}
\thanks{These two authors contributed equally}
\affiliation{Department of Electrical and Computer Engineering, Princeton University, Princeton, NJ 08544, USA}

\author{Nicholas T. Bronn}
\affiliation{IBM Quantum, IBM T.J. Watson Research Center, Yorktown Heights, NY 10598, USA}

\author{Gerasimos Angelatos}
\affiliation{Department of Electrical and Computer Engineering, Princeton University, Princeton, NJ 08544, USA}
\affiliation{RTX BBN Technologies, Cambridge, MA 02138, USA}

\author{Graham E. Rowlands}
\affiliation{RTX BBN Technologies, Cambridge, MA 02138, USA}

\author{Guilhem J. Ribeill}
\affiliation{RTX BBN Technologies, Cambridge, MA 02138, USA}

\author{Hakan E. T\"ureci}
\email{tureci@princeton.edu}
\affiliation{Department of Electrical and Computer Engineering, Princeton University, Princeton, NJ 08544, USA}

\date{\today}

\begin{abstract}
Practical implementation of many quantum algorithms known today is limited by the coherence time of the executing quantum hardware and quantum sampling noise. Here we present a machine learning algorithm, NISQRC, for qubit-based quantum systems that enables inference on temporal data over durations unconstrained by decoherence. NISQRC leverages mid-circuit measurements and deterministic reset operations to reduce circuit executions, while still maintaining an appropriate length persistent temporal memory in quantum system, confirmed through the proposed Volterra Series analysis. This enables NISQRC to overcome not only limitations imposed by finite coherence, but also information scrambling in monitored circuits and sampling noise, problems that persist even in hypothetical fault-tolerant quantum computers that have yet to be realized. To validate our approach, we consider the channel equalization task to recover test signal symbols that are subject to a distorting channel. Through simulations and experiments on a 7-qubit quantum processor we demonstrate that NISQRC can recover arbitrarily long test signals, not limited by coherence time.

\end{abstract}

\maketitle






The development of machine learning algorithms that can handle data with temporal or sequential dependencies, such as recurrent neural networks~\cite{graves_speech_2013} and transformers~\cite{vaswani2017attention}, has revolutionized fields like natural language processing~\cite{openai2023gpt}. Real-time processing of streaming data, also known as online inference, is essential for applications such as edge computing, control~\cite{canaday_modelfree_2021}, and forecasting~\cite{chattopadhyay_datadriven_2020}. The use of physical systems whose evolution naturally entails temporal correlations appears at first sight to be ideally suited for such applications. An emerging approach to learning, referred to as physical neural networks (PNNs)~\cite{wright_deep_2022, nakajima_physical_2022, markovic_physics_2020, hu_tackling_2023}, employs a wide variety of physical systems to compute a trainable transformation on an input signal. A branch of PNNs that has proven well suited to online data processing is physical reservoir computing~\cite{tanaka_recent_2019}, distinguished by its trainable component being only a linear projector acting on the observable state of the physical system~\cite{jaeger_harnessing_2004}. This approach has the enormous benefit of fast convex optimization through singular value decomposition routines and has already enabled temporal learning on various hardware platforms \cite{brunner_parallel_2013, Dong_Rafayelyan_Krzakala_Gigan_2020, canaday_modelfree_2021, rowlands_reservoir_2021, angelatos_reservoir_2021}.

Among many physical systems considered for PNNs, quantum systems are believed to offer an enormous potential for more scalable, resource-efficient, and faster machine learning~\cite{mcclean_theory_2016, havlicek_supervised_2019, Cong2019, schuld_machine_2021, cerezo_variational_2021, huang_quantum_2022, rudolph_generation_2022, wright_capacity_2019}, due to their evolution taking place in the Hilbert space that scales exponentially with the number of nodes~\cite{kalfus_hilbert_2022, mujal_opportunities_2021, fujii_harnessing_2017, chen_temporal_2020, nokkala_gaussian_2021, pfeffer_hybrid_2022, yasuda_quantum_2023}. However, quantum machine learning (QML) on present-day noisy intermediate-scale quantum (NISQ) hardware has so far been restricted to training and inference on low-dimensional static data due to several difficulties. A fundamental restriction is Quantum Sampling Noise (QSN) -- the unavoidable uncertainty arising from the finite sampling of a quantum system -- which limits the accuracy of both QML training and inference~\cite{hu_tackling_2023, garciabeni_scalable_2023, gonthier_measurements_2022} even on a fault-tolerant hardware. Additionally, the optimization landscape for training quantum systems often features ``barren plateaus" \cite{mcclean_barren_2018, wang_noise-induced_2021}, which are regions where optimization becomes exponentially difficult. These plateaus, especially in the presence of QSN, present a significant challenge to implementing QML at scales relevant to practical applications.

Two further concerns arise when considering inference on long data streams, which call into question whether quantum systems can even {\it in principle} be employed for online learning on streaming data. Firstly, without quantum error correction, the operation fidelities and finite coherence times of constituent quantum nodes places a limit on the size of data on which inference can be performed~\cite{stilck_franca_limitations_2021, dalton_variational_2022}, which would appear to rule out inference on long data streams. Secondly, the nature of measurement on quantum systems imposes a fundamental constraint on continuous information extraction over long times.
Backaction due to repeated measurements on quantum systems necessitated by inference on streaming data is expected to lead to rapid distribution of information between different parts of the system, a phenomenon known as information scrambling and thermalization~\cite{gherardini_thermalization_2021, dowling_operational_2022}, making it extremely difficult to track or retrieve the information correlations in the input data. This constraint persists even in an ideal system with perfect coherence, such as one that may be realized by a fault-tolerant quantum computer. It is not known precisely what conditions must be satisfied to avoid information scrambling. For classical dynamical systems, a strict condition known as the {\it fading memory property}~\cite{boyd_fading_1985, sandberg_series_1983} is required for a physical system to retain a persistent temporal memory that does not degrade on indefinitely long data streams. This imposes restrictions on the design of a classical reservoir and in particular how input data is encoded. Here, a mathematical framework known as Volterra Series theory~\cite{gonon_reservoir_2022} provides the basis for analyzing the memory properties of a classical dynamical system.
Such a general theory for quantum systems has remained elusive so far.

\begin{figure}[t]
    \centering
    \includegraphics[width = 1.00\columnwidth]{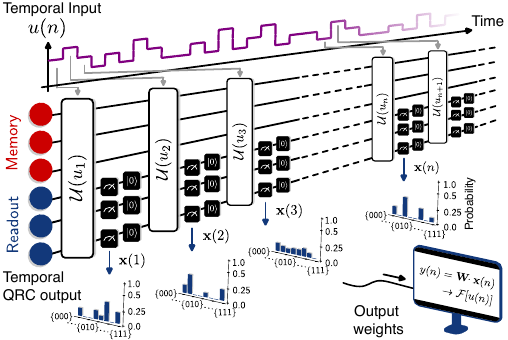}
    \caption{
    \textbf{Schematic representation of \nisqrc{} architecture for machine learning on temporal data using a convex optimization algorithm on finitely-sampled partial measurements.} For concreteness, the architecture is shown for a quantum circuit with projective computational basis readout; both the underlying quantum system and the measurement scheme can be much more general. Temporal input data is encoded into the evolution of the reservoir at every time-step $n$ via a quantum channel $\mathcal{U}(u_n)$; 
    a non-trivial I/O map is enabled via partial readout and subsequent reset of a readout subsystem, while a memory subsystem retains memory of past inputs. Temporal quantum reservoir computing (QRC) output $\mathbf{x}(n)$ are obtained via measurements (more precisely, stochastic unbiased estimators $\bar{\mathbf{X}}(n)$ of expected features are constructed from $\NS$ repetitions of the experiment, see Method \ref{app:methodFeature}), and a learned linear combination is used to approximate the target functional $y(n)$ of $u_n$.
    The overall execution time of the circuit is $O(N S)$, where $N$ is the length of input temporal sequence.
    }
    \label{fig:schematic}
\end{figure}

Here we present a Volterra theory for quantum systems that accounts for measurement backaction, necessary for analyzing the conditions required to endow a quantum system with a persistent temporal memory on streaming data. Based on this Quantum Volterra Theory we propose an algorithm, NISQ Reservoir Computing (NISQRC), that leverages recent technical advances in mid-circuit measurements to process signals of arbitrary duration,
not limited by the coherence time of constituent physical qubits (see Fig.\,\ref{fig:schematic}). The property that enables inference on an indefinitely-long input signal -- the ability to avoid measurement-induced thermalization at long times under repeated measurements due to a deterministic reset protocol -- is intrinsic to the algorithm: it survives even in the presence of QSN, and does not require operating in a precisely-defined parameter subspace -- and is thus unencumbered by barren plateaus.  

Here, we demonstrate the practical viability of NISQRC through application to a task of technological relevance for communication systems, namely, the equalization of a wireless communication channel. Channel equalization aims to reconstruct a message streamed through a noisy, non-linear and distorting communication channel and has been employed in benchmarking reservoir computing architectures~\cite{jaeger_harnessing_2004, rowlands_reservoir_2021} as well as other machine learning algorithms~\cite{burse_channel_2010, hassan_performance_2022}. This task poses a challenge for parametric circuit learning-based algorithms~\cite{schuld_machine_2021} because the number of symbols in the message, $N_{ts}$, to recover in the inference stage directly determines the length of the encoding circuit, which in turn is limited by the coherence time of the system. A more critical issue is that the recovery has to be done online, as the message is streamed, which structurally is not suitable for static encoding schemes. We demonstrate through numerical simulation (Results' subsection ``Practical machine learning using temporal data'') and experiments on a 7-qubit quantum processor (Results' subsection ``Experimental results on quantum system'') that NISQRC enables quantum systems to process signals of arbitrary duration.  Most significantly, this ability to continuously extract useful information from a single quantum circuit is not limited by coherence time.  Instead, the quantum system's coherence influences the resulting memory timescale; we show that by balancing the length of individual input encoding steps with the rate of information extraction through mid-circuit measurements, it is possible to endow the circuit with a memory that is appropriate for the ML task at hand.
Even in the limit of infinite coherence, the temporal memory is still limited by this fundamental trade-off. Reliable inference on a time-dependent signal of duration $T_\text{run} = 117 \mu\text{s}$ is demonstrated on a 7-qubit quantum processor with qubit lifetimes in the range $63 \mu\text{s}$ -- $164 \mu\text{s}$ and $T_2 = 9 \mu\text{s}$ -- $231 \mu\text{s}$. In our experiments longer durations are restricted by limitations on mid-circuit buffer clearance. To leave no doubt that a persistent memory can be generated, we first compare the experimental results to numerical simulations with the same parameters, showing excellent agreement. Building on the accuracy of numerical simulations in the presence of finite coherence and our noise model, we explicitly demonstrate successful inference on a $5000$ symbol signal: the resulting circuit duration is $500$ times that of the individual qubit lifetimes.


Here, we also develop a method to efficiently sample from deep circuits under partial measurements. Simulating individual quantum trajectories for circuits with repeated measurements requires the traversal of ever-branching paths conditioned on the measurement results, which becomes rapidly unfeasible for deep circuits. Our numerical method (see Methods' subsection ``The quantum Volterra theory and analysis of NISQRC'') allows us to sample from repeated partial measurements on circuits of arbitrary depth. We use our scheme to numerically explore other seemingly reasonable encoding methods adopted in previous studies, showing that these can lead to a sharp decline in performance when the effect of measurement is properly accounted for. Drawing upon the Quantum Volterra Theory, we unveil the underlying cause: the absence of a persistent memory mechanism. 



\section{Results}

\subsection{Time-series processing in quantum Systems}

The general aim of computation on temporal data is expressed most naturally in terms of functionals of a time-dependent input $\UI = \{u_{-\infty}, \cdots, u_{-1}, u_{0}, u_{1}, \cdots, u_{\infty} \}$. A functional $\mathcal{F}: \UI \mapsto \YO$ maps a bounded function $\UI$ to another arbitrary bounded function $\YO$, where $\YO = \{y_{-\infty}, \cdots, y_{-1}, y_{0}, y_{1}, \cdots, y_{\infty} \}$. Without loss of generality these functions can be normalized; we choose $u_n \in [-1,1]$ and $y_n \in [-1,1]$. 
Within the reservoir computing paradigm~\cite{nakajima_fischer_book_2021}, this processing is achieved by extracting outputs $\mathbf{x}(n)$, where $n$ is a temporal index, from a physical system evolving under said time-dependent stimulus $u_n \equiv \UI(n)$. Learning then entails finding a set of optimal time-independent weights $\mathbf{w}$ to best approximate a desired $\mathcal{F}$ with a linear projector $y_n \equiv \YO(n) = \mathbf{w} \cdot \mathbf{x}(n)$. 
If the physical system is sufficiently complex, its temporal response $\mathbf{x}(n)$ to a time-dependent stimulus $\mathbf{u}$ is universal in that it can be used to approximate a large set of functionals $\mathcal{F}[\mathbf{u}]$ with an error scaling inversely in system size and using only this simple linear output layer~\cite{ gonon_reservoir_2020, chen_temporal_2020, nokkala_gaussian_2021}. 

To analyze the utility of this learning framework, it proves useful to quantify the space of functionals $\mathcal{F}[\mathbf{u}]$ that are accessible. For classical non-linear systems, a firmly-established means of doing so is a Volterra series representation of the input-output (I/O) map~\cite{boyd_fading_1985}:
\begin{align}
    x_j (n) = \sum_{k = 0}^{\infty} \sum_{n_1 = 0}^{\infty} \cdots \hspace{-3mm} \sum_{n_k = n_{k-1}}^{\infty} \hspace{-3mm} h_k^{(j)} (n_1, \cdots, n_k) \prod_{\kappa = 1}^{k} u_{n - n_{\kappa}} \label{eq:Volterra}
\end{align}
where the Volterra kernels $h_k^{(j)} (n_1, \cdots, n_k)$ characterize the dependence of the systems' measured output features at time $n$ on its past inputs $u_{n - n_{\kappa}}$. Hence the support of $h_k^{(j)}$ over the the temporal domain $(n_1,\cdots,n_k)$ quantifies the notion of {\it memory} of a particular physical system, with the kernel order $k$ being the corresponding degree of nonlinearity of the map. Most importantly, the Volterra series representation describes a time-invariant I/O map, as well as the property of fading memory, which roughly translates to the property that the reservoir forgets initial conditions and thus depends more strongly on more recent inputs~(for instance, for multi-stable dynamical systems, a global representation such as \Eq{eq:Volterra} may not exist. However a local representation around each steady state can be shown to exist with a finite convergence radius).
The realization of such a time-invariant map is essential for a physical system to be reliably employed for inference on an input signal of arbitrary length, and thus for online time series processing.

In classical physical systems, the existence of a unique information steady state and the resulting fading memory property is determined only by the input encoding dynamics -- the map from input series to system state.
More explicitly, the information extraction step (sometimes referred to as the ``output layer") on a classical system is considered to be a passive action, so that the state can always be observed at the precision required. However, for physical systems operating in the quantum regime, the role of quantum measurement is fundamental: in addition to the inherent uncertainty in quantum measurements as dictated by the Heisenberg uncertainty principle, the conditional dependence of the statistical system state on prior measurement outcomes -- referred to as backaction -- strongly determines the information that can be extracted. Recent work in circuit-based quantum computation has shown that the qualitative features of the statistical steady state of monitored circuits strongly depends on the rate of measurement \cite{Skinner2019, block_measurementinduced_2022}. 
In particular, generic quantum systems that alternate dynamics and measurement (input encoding and output in the present context) are known to give rise to deep thermalization of the memory subsystem~\cite{choi_preparing_2023, ippoliti_dynamical_2023}, resulting in an approximate Haar-random state with vanishing temporal memory. 
The absence of a comprehensive framework in QML for analyzing and implementing an encoding-decoding system with finite temporal memory, along with characterization tools for the accessible set of input-output functionals, has hindered both a systematic study and the practical application of online learning methods.

Here, we develop both a general temporal learning framework suitable for qubit-based quantum processors and the associated methods of analysis based on an appropriate generalization of the Volterra Series analysis to monitored quantum systems, the Quantum Volterra Theory (QVT). Our approach incorporates the effects of backaction that results from quantum measurements in the process of information extraction. 

We begin by providing a fundamental description of both the information input and output processes that enable general time series processing with quantum systems, before specializing to the \nisqrc{} algorithm. The `input' component of the map is given by a pipeline (encoding) that injects temporal data $\{ u_n \}$ into a quantum system through a general parameterized quantum channel $\mathcal{U}(u_n)\hat{\rho}$
.  This channel could describe for instance continuous Linblad evolution for a duration $\tau$, namely $e^{\tau\mathcal{L}(u_n)}\hat{\rho}$, as in Results' subsection ``Practical machine learning using temporal data'', or a discrete set of gates as in Results' subsection ``Experimental results on quantum system''; $\mathcal{U}(u_n)$ is generally applied to all qubits, and we assume only that they are not explicitly monitored for its duration.
 
To enable persistent memory in the presence of quantum measurement, we separate the $L$-qubit system into $M$ memory qubits and $R$ readout qubits ($L=M+R$), and denote their respective Hilbert spaces with superscript \textsf{M} and \textsf{R}. After evolution under any input $u_n$, only the $R$ readout qubits are (simultaneously) measured; this separation therefore allows for the concept of partial measurements of the full quantum system, which proves critical to the success of NISQRC. The measurement scheme itself can be very general, characterized by a positive operator-valued measure (POVM) 
\begin{equation}
\mathcal{O}_{R} = \left\{ \hat{M}_j \left| \hat{M}_j = \hat{I}^{\otimes M} \otimes \hat{E}_j \right. \right\}
\label{eq:nisqrcPOVM}
\end{equation}
satisfying $\hat{E}_j \succeq 0$ and $\sum_j \hat{E}_j = \hat{I}^{\otimes R}$.
Here we will consider a practically implementable measurement in the readout qubit computational basis,  described by $\hat{E}_j = \ket{\mathbf{b}_j} \! \bra{\mathbf{b}_j}$:
each bit-string $\mathbf{b}_j$ is the $R$-bit binary representation of integer $j \in \{0, 1, \cdots, 2^R-1\}$ denoting the bit-wise state of the measured qubits.

As elucidated by the QVT analysis of Results' subsection ``Quantum Volterra Theory'', a purification mechanism must necessarily accompany readout to prevent thermalization and furnish our quantum architecture with persistent fading memory.  This is accomplished by following each projective measurement operation with a deterministic reset to the ground state $\ket{0}$.  The resulting measure-reset operation we employ throughout this paper is formally described by the POVM operators $\hat{E}_j = \hat{K}_j^{\dagger}\hat{K}_j$ in \Eq{eq:nisqrcPOVM}, with non-hermitian Kraus operators $\hat{K}_j = \ket{\mathbf{b}_0}\!\bra{\mathbf{b}_j}$. In each measure-reset step, only the readout qubits are measured in the computational basis and then reset to the ground state, irrespective of the measurement outcome.

\nisqrc{} is distinguished by the iterative encode-measure-reset scheme depicted in Fig.~\ref{fig:schematic}. Explicitly, for a given input sequence $\mathbf{u}$ with length $N$, we initialize the system in the state $\hat{\rho}^{\mathsf{M}}_0  \!\otimes\! \ket{0}\!\bra{0}^{\otimes R}$. For each element of the input sequence $u_n$, an encoding step is comprised of unmonitored evolution of all qubits via $\mathcal{U}(u_n)$ followed by a measure-reset operation $\mathcal{O}_{R}$.  The measurement outcome in this single shot is a random bitstring $\mathbf{b}^{(s)}(n)$, and the resulting state is $\hat{\rho}^{\mathsf{M},\mathtt{cond}}_n  \!\otimes\! \ket{0}\!\bra{0}^{\otimes R}$: the memory qubits are in a state conditioned on the measurement outcome, and the readout qubits are reset.  The subsequent input is then encoded in this state, i.e.~$\mathcal{U}(u_{n+1})\,\big(\hat{\rho}^{\mathsf{M},\mathtt{cond}}_n \!\otimes\!\ket{0}\!\bra{0}^{\otimes R}\big)$, and the process is iterated as long as there is data in the pipeline. This structure elucidates the naming of the unmeasured memory qubits: these are the only qubits that retain memory of past inputs.

The above description yields a set of $N$ measurement outcomes $\{\mathbf{b}^{(s)}(n)\}$ observed in a single shot $s$ of the quantum circuit.  In order to obtain statistics and therefore to output features as expected values of observables $\hat{M}_j$, we perform $S$ repetitions of this circuit for a given $\mathbf{u}$ sequence: the total execution time is $NS$, linear with respect to shots $S$ and input length $N$. The resulting readout features are formally defined as the probability
\begin{align}
    x_j(n) = \mathrm{Pr}[\mathbf{b}^{(s)}(n) = \mathbf{b}_j | \mathbf{u}], 
\end{align}
which are estimated by the empirical mean $\bar{X}_j(n) = (1/S) \sum_{s} \delta(\mathbf{b}^{(s)}(n), \mathbf{b}_j)$ of $\{\mathbf{b}^{(s)}(n)\}_{s\in[S]}$ (see Methods' subsection ``Generating features via conditional evolution and measurement'' for more details of NISQRC algorithm). We show in Supplementary Note 2 that at time step $n$, $x_j(n)$ can be computed efficiently through
\begin{align}
    x_j(n) = \mathrm{Tr} ( \hat{M}_j \hat{\rho}^{\mathsf{MR}}_n),
    \label{eq:outputFeature}
\end{align}
where $\hat{\rho}^{\mathsf{MR}}_n$ is the effective full $L$-qubit system state at time step $n$ prior to measurement.

The output $y_n \equiv \YO(n) = \mathbf{w} \cdot \mathbf{x}(n) \in \mathbb{R}$ is obtained from the measurement results in each step, defining the functional I/O map which we characterize next (see details in Methods' subsection ``Generating features via conditional evolution and measurement'' and ``The quantum Volterra theory and analysis of NISQRC''). 
This complete architecture, from the quantum circuit generating measurement outcomes for a given input, to the construction of weighted output features, is depicted schematically in Fig.~\ref{fig:schematic}.
We note that reset operations have been used implicitly in prior work on quantum reservoir algorithms, where the successive inputs are encoded in the state of an `input' qubit~\cite{fujii_harnessing_2017, mujal_time-series_2023}. However the critical role of the reset operation in endowing a quantum reservoir with a persistent memory, discussed in the next section, has so far not been highlighted.

While for null inputs (i.e.~$u_n=0$ for all $n$) such quantum systems are guaranteed to have a unique statistical steady state, the existence of a nontrivial memory and kernel structure is much more involved. Through QVT (see Methods' subsection ``The quantum Volterra theory and analysis of NISQRC''), we show that these requirements place strong constraints on the encoding and measurement steps viz.~the choice of ($\mathcal{U}$, $\hat{M}_j$). This then enables us to propose an algorithm for online learning that provably provides a controllable and time-invariant temporal memory (which will be referred to as {\it persistent memory}) -- enabling inference on arbitrarily long input sequences even on NISQ hardware without any error-mitigation or correction.

\subsection{Quantum Volterra theory}
\label{sec:theory}

In \nisqrc{} the purpose of the partial reset operation is to endow the system with asymptotic time-invariance, a finite persistent memory and a nontrivial Volterra Series expansion for the system state (see Methods' subsection ``The quantum Volterra theory and analysis of NISQRC'' and Supplementary Note 3):
\begin{align}
    \hat{\rho}^{\mathsf{MR}}_n = \sum_{k = 0}^{\infty} \sum_{n_1 = 0}^{\infty} \cdots \hspace{-3mm} \sum_{n_k = n_{k-1}}^{\infty} \hspace{-3mm} \hat{h}_k (n_1, \cdots, n_k) \prod_{\kappa = 1}^{k} u_{n - n_{\kappa}}. \label{eq:QVT}
\end{align} 
where all Volterra kernels $\hat{h}_k$ are quantum operators. The classical kernels in Eq.\,(\ref{eq:Volterra}) describing the {\it measured features} can be extracted through $h^{(j)}_k = \mathrm{Tr}(\hat{M}_j \hat{h}_k)$. We refer to this analysis as the Quantum Volterra Theory (QVT). Through analytical arguments based on the QVT, we show that omitting the partial reset operation renders all Volterra kernels trivial -- a finding corroborated by our experimental results in Results' subsection ``Experimental results on quantum system''. 

QVT also provides a way to characterize the important memory time-scales of the I/O map generated by the \nisqrc{} algorithm through a given encoding, which we use in Results' subsection ``Practical machine learning using temporal data'' to aid encoding design for a specific ML task on an experimental system. In what follows, we show that inference on an indefinitely long input sequence can be done even in the presence of dissipation and decoherence. 

Consider an input-encoding $\mathcal{U}(u_n)\hat{\rho} = e^{\tau\mathcal{L}(u_n)}\hat{\rho}$ where 
\begin{equation}
    \mathcal{L} (u) \hat{\rho} = - i [\hat{H} (u), \hat{\rho}] +\mathcal{D}_\mathrm{T} \hat{\rho},
    \label{eq:dissipation}
\end{equation}
representing evolution under a parameterized Hamiltonian $\hat{H}(u_n)$ for a duration $\tau$ in the presence of dissipation $\mathcal{D}_\mathrm{T}$.  For concreteness, we take  $\mathcal{D}_\mathrm{T} = \sum_{i = 1}^L \gamma_i \mathcal{D} [\hat{\sigma}_i^{-, z}]$ describing decoherence processes and study here a specific Ising Hamiltonian encoding $\hat{H}(u) = \hat{H}_0 + u\cdot\hat{H}_1$ inspired by quantum annealing and simulation architectures (other ans\"atze can likewise be considered), 
\begin{align}
    \hat{H}_0 = \sum_{\langle i, i'\rangle} J_{i,i'} \hat{\sigma}^z_i \hat{\sigma}^z_{i'} + \sum^L_{i=1} \eta^x_{i} \hat{\sigma}^x_i, ~
    \hat{H}_1 = \sum^L_{i=1} \eta^z_{i} \hat{\sigma}^z_i.
    \label{eq:H0_H1}
\end{align}
The coupling strength $J_{i,i'}$, transverse $x$-field strength $\eta^x_{i}$ and longitudinal $z$-drive strength $\eta^z_{i}$ are randomly chosen, but then fixed for all inputs $\{u_n\}$ (see Supplementary Note 1 for more details).  The encoding channel is applied for duration $\tau$, and each qubit has a finite lifetime $T_1 =  \gamma^{-1}$. We will specify the number of memory and readout qubits of a given QRC with the notation $(M+R)$.


\begin{figure}[t]
    \centering
    \includegraphics[width = 1.00\columnwidth]{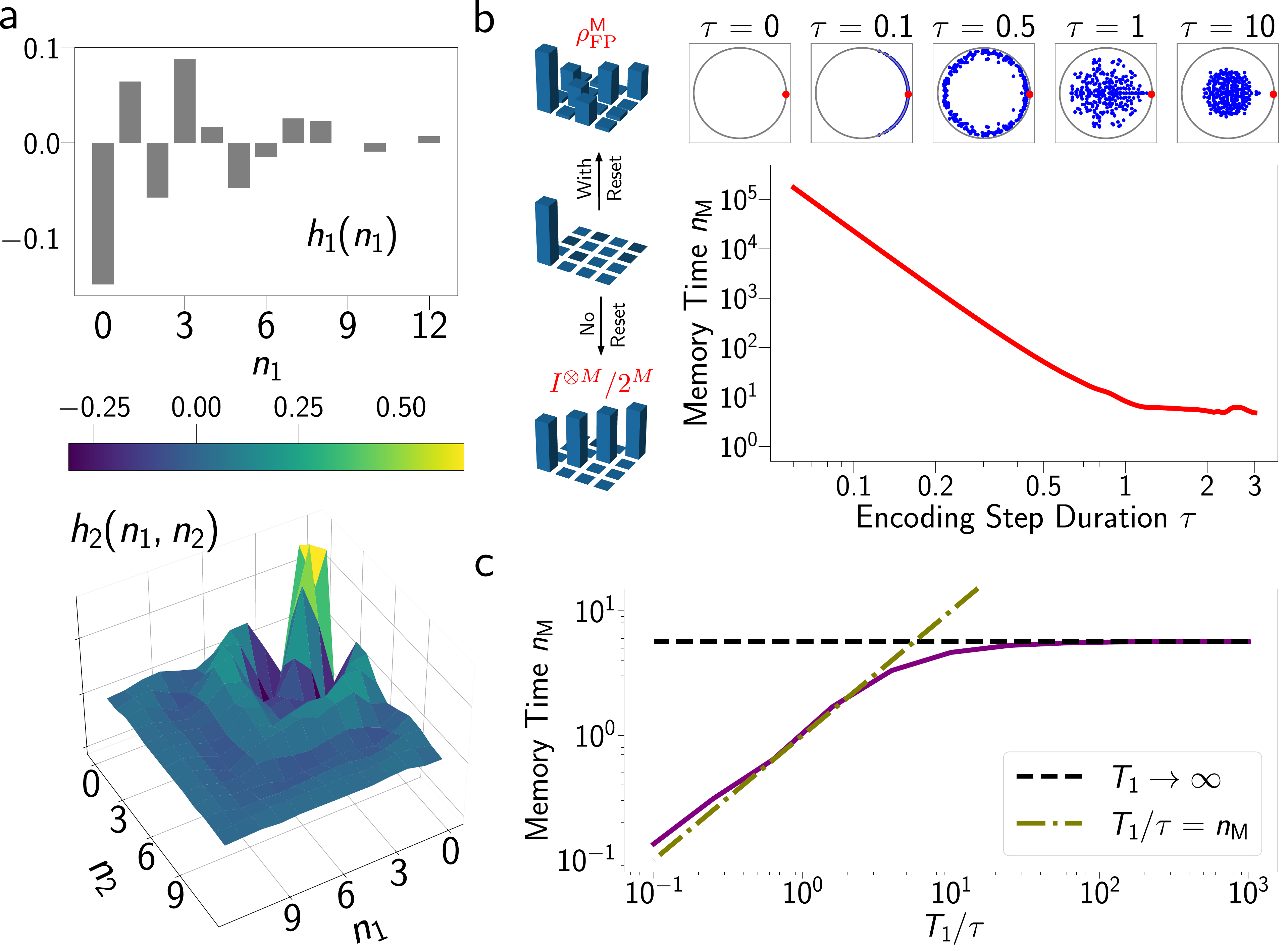}
    \caption{\textbf{Quantum Volterra Theory (QVT) analysis for $(M+R)$-qubit reservoir.} 
    (a) First and second order Volterra kernels in a $(2+1)$-qubit QRC, which vanish at large $n_1$ and $n_2$ due to finite memory $n_{\rm M}$.
    (b) Fixed-point of memory subsystem $\hat{\rho}^{\mathsf{M}}_{\rm FP}$ with reset (top) and without reset (bottom), starting from an arbitrary initial state (center). Without reset, the fixed point is always the trivial fully-mixed state and Volterra kernels vanish.
    Top panel shows the distribution of the $4^M=256$ eigenvalues of $\mathcal{P}_0$ in a $(4+2)$-qubit QRC, where red dots correspond to the static unit eigenvalue $\lambda_1=1$. 
    The remaining eigenvalues $\lambda_{\alpha \geq 2}$ (blue) evolve with evolution time $\tau$, leading to a variable memory time. Bottom panel shows the resulting memory time $n_{\mathrm{M}}$ as a function of the evolution duration $\tau$. (c) Memory time $n_{\mathrm{M}}$ as a function of qubit lifetimes $T_1=\gamma^{-1}$, in terms of the evolution duration $\tau$ in a $(4+2)$-qubit QRC. Provided $T_1 \gg \tau$, $n_{\rm M} \to n_{\rm M}^0$, so that the QRC memory is mostly dominated by its lossless dynamical map, and not by $T_1$ in this regime.
    }
    \label{fig:nM_T1}
\end{figure}


In Fig.~\ref{fig:nM_T1}(a) we plot the first two Volterra kernels $h_1$ and $h_2$ (cf.~\Eq{eq:Volterra}) for a random $(2+1)$-qubit QRC using the above encoding and the reset scheme. The expression for these kernels have been derived from the QVT; their numerical construction is discussed in Methods, also see Supplementary Equations 43-45. Importantly, we find all kernels have an essential dependence on the statistical steady state or fixed-point in the absence of any input: $\hat{\rho}^{\mathsf{M}}_{\rm FP} = \lim_{n\to\infty} \hat{\rho}^{\mathsf{M}}_n \big|_{u_n = 0}$. Here $\hat{\rho}^{\mathsf{M}}_n \big|_{u_n = 0}= \mathcal{P}_0^n\hat{\rho}^{\mathsf{M}}_0$ is obtained by $n$ applications of the null-input single-step quantum channel $\mathcal{P}_0$, defined in Methods' subsection ``The quantum Volterra theory and analysis of NISQRC''. 
The properties of quantum Volterra kernels, including their characteristic decay time, can be related to the spectrum of $\mathcal{P}_0$, defined by $\mathcal{P}_0 \hat{\varrho}^{\mathsf{M}}_\alpha = \lambda_\alpha \hat{\varrho}^{\mathsf{M}}_\alpha$.
Here $\hat{\varrho}^{\mathsf{M}}_\alpha$ are eigenvectors that exist in the $4^M$-dimensional space of memory subsystem states. The eigenvalues satisfy $1 = \lambda_1 \geq |\lambda_2| \geq \cdots \geq |\lambda_{4^M}| \geq 0$; examples are plotted in Fig.~\ref{fig:nM_T1}(b) for various values of $\tau$. The unique eigenvector corresponding to the largest eigenvalue $\lambda_1 = 1$ is special, being the fixed-point of the memory subsystem, $\hat{\varrho}^{\mathsf{M}}_1 = \rhofpm$, reached once transients have died out.

The second largest eigenvalue $\lambda_2$ determines the time over which memory of an initial state persists as this fixed point is approached, and is used to identify a {\it memory time} $n_\mathrm{M} = - 1/\ln|\lambda_{2}|$. Note that this quantity is dimensionless and can be converted to actual passage of time through multiplication by $\tau$, while $n_\mathrm{M}$ itself non-trivially depends on $\tau$ (see Fig.~\ref{fig:nM_T1}(b)). The memory time describes an effective `envelope' for a system's Volterra kernels; additional nontrivial structure is also required for QRC to produce meaningful functionals of past inputs. With the spectral problem at hand, we next analyze the information-theoretical benefit of the reset operation. Firstly, the absence of the unconditional reset operation produces a \textit{unital} $\mathcal{P}_0$~(``unital'' refers to an operator that maps the identity matrix to itself). with resulting $\rhofpm = I^{\otimes M}/2^M$. This fully-mixed state is inexorably approached after $n_{\mathrm{M}}$ steps under any input sequence and retains no information on past inputs: all Volterra kernels therefore vanish, despite a generally-finite $n_{\rm M}$. 
Such algorithms (e.g.~Refs.\,\cite{yasuda_quantum_2023}) are only capable of processing input sequences of length $n_\mathrm{M}$ and would not retain a persistent memory necessary for inference on longer sequences of inputs.
Hence such encodings would be unsuitable for online learning on streaming data. The possibility of inference through the transients have been observed and utilized before (see e.g.~Ref.~\cite{larger_photonic_2012, brunner_parallel_2013, fan_learning_2022}) in the context of classical reservoir computing. However, the simple yet essential inclusion of the purifying reset operation avoids unitality -- more generally, a common fixed point for all $u$-encoding channels -- which we find is the key to enabling nontrivial Volterra kernels and consequent online QRC processing (see Methods' subsection ``The quantum Volterra theory and analysis of NISQRC'' and also Ref.\,\cite{martinez-pena_quantum_2023}). 
Once such an I/O map is realized, $\lambda_{\alpha}$ and the consequent memory properties can be meaningfully controlled by the QRC encoding parameters. As shown in Fig.~\ref{fig:nM_T1}(b) the characteristic decay time set by $n_{\rm M}$, for instance, decreases across several orders-of-magnitude with increasing $\tau$.

The partial measurement and reset protocol also resolves the unfavorable quadratic runtime scaling of prior approaches. A wide range of proposals and implementations of QRC~\cite{suzuki_natural_2022, chen_temporal_2020, pfeffer_hybrid_2022} consider the read out of all constituent qubits at every output step, terminating the computation. Not only does this preclude inference on streaming data, it requires the entire input sequence to be re-encoded to proceed one step further in the computation, leading to an $O(N^2 S)$ running time. As shown in schematic Fig.\,\ref{fig:schematic}, incorporating partial measurement with reset in \nisqrc{} does not require such a re-encoding; the entire input sequence can be processed in any given measurement shot $\NS$, enabling online processing with an $O(N S)$ runtime, while maintaining a controllable memory timescale. We note that an alternative scheme to remedy this issue has been suggested in Ref.~\cite{mujal_time-series_2023}, which relies on information extraction through continuous weak measurement.

Next we show that the nontrivial nature of Volterra kernels realized by the \nisqrc{} algorithm is preserved under the inclusion of dissipation. For example, we explore the effect of finite qubit $T_1$ on $n_{\rm M}$ in Fig.~\ref{fig:nM_T1}(c). If $T_1/\tau > n_{\rm M}^0$, where $n_{\rm M}^0$ is the memory time of the lossless map, then $n_{\rm M} \to n_{\rm M}^0$ and is essentially independent of $T_1$, determined instead by the unitary and measurement-induced dynamics. Therefore the design of the encoding algorithm has to be guided by matching the memory time of the reservoir to the longest correlation time in the input data. Additional design criteria are discussed in Section Discussion. As a result, lossy QRCs can still be deployed for online processing, with a total run time $T_{\rm run}$ that is unconstrained by (and can therefore far exceed) $T_1$. We will demonstrate this via simulations in Results' subsection ``Practical machine learning using temporal data'' with $T_{\rm run} \gg T_1$, and via experiments in Results' subsection ``Experimental results on quantum system'' for $T_{\rm run} \simeq T_1$; in the latter $T_{\rm run}$ is limited only by memory buffer constraints on the classical backend.

\subsection{Practical machine learning using temporal data}
\label{sec:mlexample}

Thus far, we have assumed outputs to be expected features $x_j(n)$, which in principle assumes an infinite number of measurements. In any practical implementation, one must instead estimate these features with $S$ shots or repetitions of the algorithm for a given input $\UI$.  The resulting QSN constrains the learning performance achievable in experiments on quantum processors in a way that can be fully characterized~\cite{hu_tackling_2023}, and is therefore also included in numerical simulations which we present next.

To demonstrate the utility of the \nisqrc{} framework, we consider a practical application of machine learning on time-dependent classical data: the \textit{channel equalization} (CE) task~\cite{jaeger_harnessing_2004, rowlands_reservoir_2021}. Suppose one wishes to transmit a message $m(n)$ of length $N$, which here takes discrete values from $\{-3, -1, 1, 3\}$, through an unknown noisy channel to a receiver. This medium generally distorts the signal, so the received version $u(n)$ is different from the intended $m(n)$. Channel equalization seeks to reconstruct the original message $m(n)$ from the corrupted signal $u(n)$ as accurately as possible,  and is of fundamental importance in communication systems. Specifically, we assume the message is corrupted by nonlinear receiver saturation, inter-symbol interference (a linear kernel), and additive white noise \cite{jaeger_harnessing_2004, rowlands_reservoir_2021} (additional details in Supplementary Note 6). As shown in Fig.~\ref{fig:CE_test_error}(a), even if one has access to the exact inverse of the resulting nonlinear filter, the signal-to-noise (SNR) of the additive noise bounds the minimum achievable error rate.  We also show the error rates of simply rounding $u(n)$ to the nearest $m$, and a direct logistic regression on $u(n)$ (i.e.~a single-layer perceptron with a softmax activation -- see Supplementary Note 6).
for comparison.  Both these approaches are linear and memory-less and therefore perform poorly on the non-trivial nonlinear filter we consider, although logistic regression outperforms rounding ($\approx$$30\%$) by inverting the linear portion of the distortion.


\begin{figure}[t]
    \centering
    \includegraphics[width = 1.00\columnwidth]{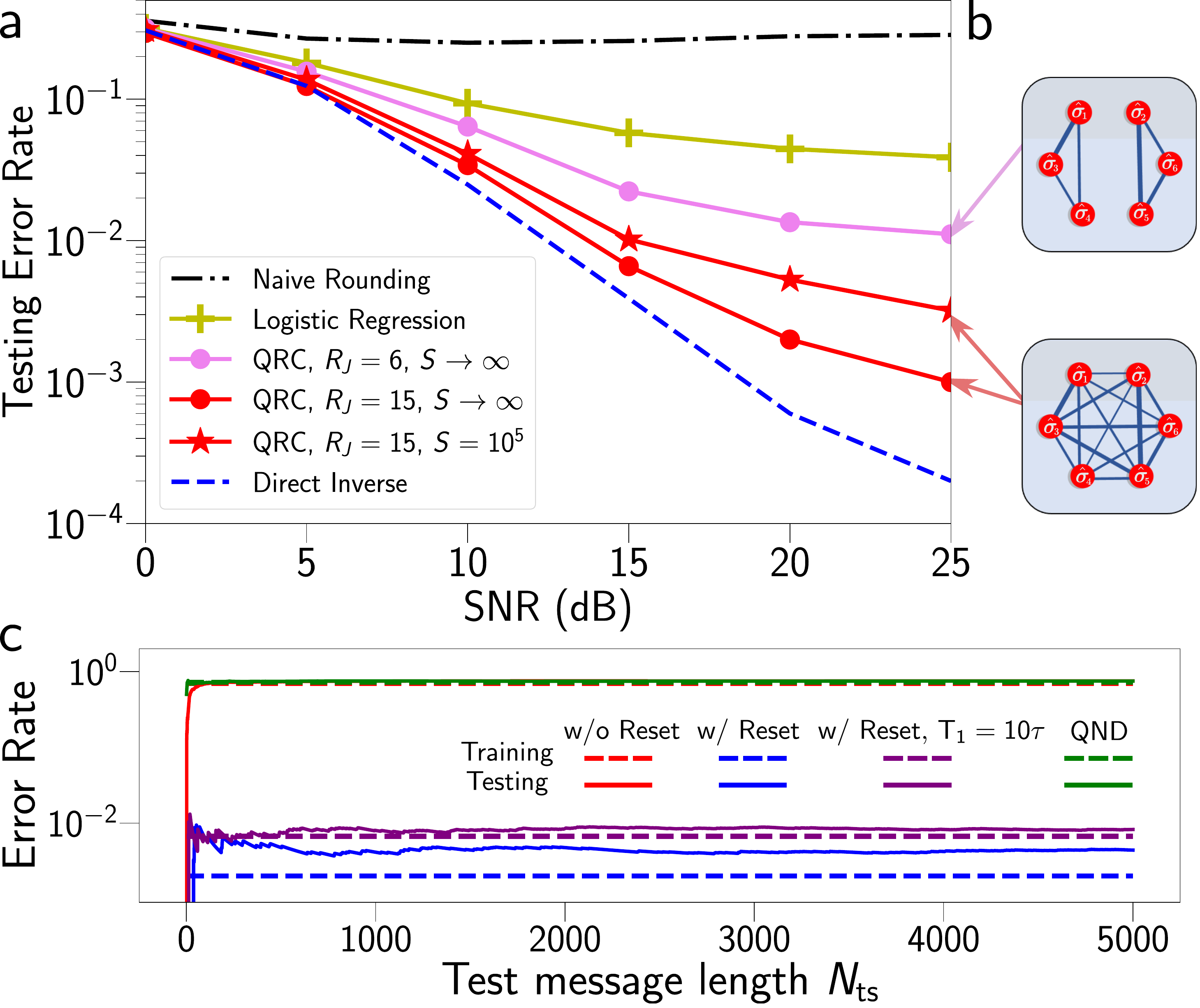}
    \caption{\textbf{Numerical results for the channel equalization (CE) task with Hamiltonian ansatz.} (a) Error rates on test messages for the CE task with a Hamiltonian ansatz $(2+4)$-qubit QRC for two distinct connectivities shown in (b) The fully-connected QRC in red has Jacobian rank $R_J = 2^R-1=15$ and is shown for both $\NS\to\infty$ (circles) and finite $\NS=10^5$ ($\star$), whereas the split QRC has $R_J = 2(2^2-1)=6$ and only $\NS\to\infty$ is plotted in magenta.  These are compared with the error rates of naive rounding (black dash-dots) and logistic regression on the current signal (yellow $+$, see Supplementary Note 6), and the exact channel inverse (blue dashed).
    (c) Performance of connected QRC on ${\rm SNR}=20~$dB test signals (solid) of increasing length $N_{ts}\leq 5000$, with shots $S=10^5$.  Training error on $N=100$-length messages is indicated for comparison in dashed lines. Without reset (red) or using 4 ancilla qubit ansatz with quantum non-demolition (QND) readout (proposed in Ref.\,\cite{yasuda_quantum_2023}, green), the algorithms both fail, approaching the random guessing error rate and showing that both architectures suffer from the thermalization problem. Performance is only slightly reduced from the dissipation-free case (blue) when strong decay $T_1 = 10 \tau$ is included (purple). 
    All error rates in (c) are averaged over $8$ different test messages.
    }
    \label{fig:CE_test_error}
\end{figure}


We now perform the CE task using the \nisqrc{} algorithm on a simulated $(2+4)$-qubit reservoir under the ansatz of Eq.~(\ref{eq:H0_H1}), as could be realized in quantum annealing hardware (see Fig.~\ref{fig:CE_test_error}). We will later demonstrate the same task in experiments with a completely different quantum system and encoding ansatz, implemented on a superconducting quantum processor (see Fig.~\ref{fig:CE_expt_error}). The ability to efficiently compute the Volterra kernels for this quantum system immediately provides guidance regarding parameter choices. In particular, we choose random parameter distributions such that the average (across the circuit) $J_{i,i'}\tau$, $\eta^x_{i}\tau$ and $\eta^z_{i}\tau$ provides a memory time $n_{\mathrm{M}} \approx O(10^1)$, on the order of the length of the distorting linear kernel $h(n)$, which is $8$. These QRCs have $K=2^4=16$ readout features $\{x_j(n)\}_{j \in [K]}$ whose corresponding time-independent output weights $\mathbf{w}$ are learned by minimizing cross-entropy loss on $100$ training messages of length $N=100$ (see Supplementary Note 6 for additional details). The resulting \nisqrc{} performance on test messages is studied in Fig.\,\ref{fig:CE_test_error}(a), where we compare two distinct coupling maps shown in Fig.\,\ref{fig:CE_test_error}(b).  In the highly-connected (lower) system the performance approaches the theoretical bound for $\NS\to\infty$; finite sampling  (here, $\NS=10^5$ is in the range typically used in experiments) increases the error rate as expected, but the increase in error rate in numerical simulations is observed to depend on the encoding (not reported here). In all cases, \nisqrc{} significantly outperforms direct logistic regression due to its ability to reliably implement non-linear memory kernels and therefore approximate the distorting channel inverse.

We note that the split system (upper) performs significantly worse even without sampling noise: this is because the quantum system lives in a smaller effective Hilbert space -- the product of two disconnected three-qubit systems -- and is far less expressive as a result.  Although in both cases the number of measured features is the same, those from the connected system span a richer and independent space of functionals.  This functional independence can be quantified by the Jacobian rank $R_J$, which is the number of independent $\UI$-gradients that can be represented by a given encoding (Supplementary Note 5); an increased connectivity and complexity of state-description generally manifests as an increase in the Jacobian rank and consequent improved CE task performance. This observation can be viewed as a generalization of the findings in time-independent computation~\cite{hu_tackling_2023} to tasks over temporally-varying data, and also agrees with related recent theoretical work~\cite{pfeffer_hybrid_2022}. 

Most importantly, we demonstrate in Fig.~\ref{fig:CE_test_error}(c) that the \nisqrc{} algorithm enables the use of a quantum reservoir for online learning. In all cases studied here, $N=100$ is used for training and the length of the $\rm{SNR}=20~$dB test messages $N_{ts}$ is varied. As suggested by the QVT, the performance is unaffected by $N_{ts}$ even if it greatly exceeds the lifetime of individual qubits: $N_{ts} = T_{\rm run}/\tau \gg T_1/\tau = 10$, and \nisqrc{} can therefore be used to perform inference on an indefinite-length signal with noisy quantum hardware.  As seen in the same figure, while dissipation imposes only a small constant performance penalty, the reset operation is critical: if removed, the error rate increases to that of random guessing, as the Volterra kernels vanish and the I/O map becomes trivial.

We finally note that an arbitrarily-inserted reset operation may not be sufficient to create a non-zero persistent memory. For instance, an analysis based on the QVT shows that despite its use in a recently studied reservoir algorithm~\cite{yasuda_quantum_2023} (based on a quantum non-demolition measurement proposal in Ref.\,\cite{chen_temporal_2020}), the reset operation can not avoid a zero persistent memory, effectively resulting in an amnesiac reservoir. In this scheme, the quantum circuit is coupled to ancilla qubits by using transversal CNOT gates. Upon closer examination it is found that while the projective measurement of ancilla qubits leads to read out of system qubits and their collapse to the ancilla state via backaction, subsequent reset of the ancillas does not reset the system qubits. This scheme therefore suffers from the same thermalization problem as any no-reset \nisqrc{} does, and hence has zero persistent memory. We verify this analysis in Fig.\,\ref{fig:CE_test_error}(c) by implementing the CE task with a four-ancilla-qubit circuit. The error rates are found to be very close to the no-reset-\nisqrc{} one, whose I/O map we have shown before to be trivial (see also Fig.\,\ref{fig:CE_test_error}(c)).

\subsection{Experimental results on quantum systems}
\label{sec:NISQ_implementability}

\begin{figure}[t]
    \centering
    \includegraphics[width = 1.00\columnwidth]{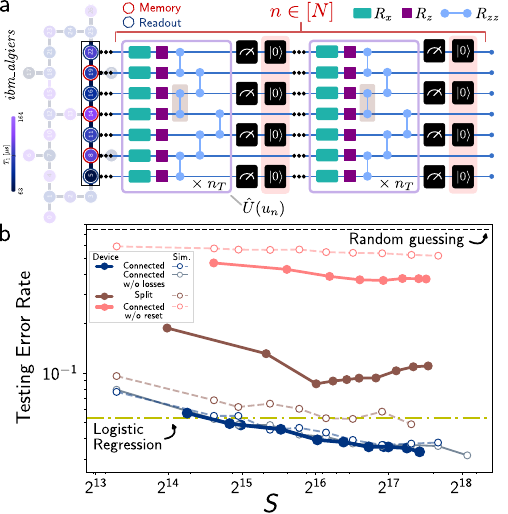}
    \caption{\textbf{Experimental results for the channel equalization task with circuit ansatz.} (a) $(3+4)$-qubit linear chain of the \textit{ibm\_algiers} device used to perform the CE task. Filled colors represent qubit $T_1$ time according to the displayed colorbar, for the specific experimental run with the split chain. Qubits indexed \{8, 14, 19\} are used for memory and qubits \{5, 11, 16, 22\} for readout, and gate-decomposition of the encoding unitary $\hat{U}(u_n)$ is depicted. Removing gates shaded in brown yields two smaller chains to explore the role of connectivity, while removing reset operations (shaded peach) allows switching from a non-unital to a unital I/O map. (b) Testing error rates for the $\rm{SNR}=20~$dB CE task of Results' subsection ``Practical machine learning using temporal data'' with $N=20$ on the \textit{ibm\_algiers} device in filled circles and in simulation in open circles, as a function of number of shots $\NS$.  The connected circuit in blue outperforms the split circuit in brown and the circuit without reset in peach.  For comparison, we plot the testing error rate of logistic regression (yellow line), as well as random guessing (black dashed line).
    }
    \label{fig:CE_expt_error}
\end{figure}

We now demonstrate \nisqrc{} in action by performing the $\rm{SNR}=20~$dB CE task on an IBM Quantum superconducting processor. To highlight the generality of our \nisqrc{} approach, we now consider a circuit-based parametric encoding scheme inspired by a Trotterization of Eq.~(\ref{eq:H0_H1}), suitable for gate-based quantum computers. In particular, we use a $L=7$ qubit linear subgraph of the \textit{ibm\_algiers} device, with $M=3$ memory qubits and $R=4$ readout qubits in alternating positions, as depicted in Fig.\,\ref{fig:CE_expt_error}(a). The encoding unitary for each time step $n$ is also shown: $\hat{U}(u_n) = \left(\mathcal{W}(J)  \mathcal{R}_{z}(\bm{\theta}^z +\bm{\theta}^I u_n ) \mathcal{R}_{x}(\bm{\theta}^x) \right)^{n_{T}}$, where $\mathcal{R}_{x,z}$ are composite Pauli-rotations applied qubit-wise, and $\mathcal{W}(J)$ defines composite $\mathcal{R}_{zz}$ gates between neighbouring qubits, all repeated $n_T=3$ times (for parameters $\bm{\theta}^{x,z,I}, J$ and further details see Methods' subsection ``IBM Quantum implementation'').


Realizing the \nisqrc{} framework with the circuit ansatz depicted in Fig.~\ref{fig:CE_expt_error}(a) requires the state-of-the-art implementation of mid-circuit measurements and qubit reset, which has recently become possible on IBM Quantum hardware~\cite{hua_exploiting_2022}. We plot the testing error using the indicated linear chain of the \textit{ibm\_algiers} device as a function of the number of shots $\NS$ in solid blue Fig.~\ref{fig:CE_expt_error}(b), alongside simulations of both the ideal unitary circuit and with qubit losses in open circles. We clearly observe that performance is influenced by the number of shots available, and hence by QSN. In particular, for a sufficiently large $S$, the device outperforms the same logistic regression method considered previously. For the circuit runs, the average qubit coherence times over $7$~qubits are $T_1^\text{av} = 124~\mu$s, $T_2^\text{av} = 91\mu$s (see Supplementary Note 9 for the ranges of all parameters, which varies over the time of runs as well), while the total circuit run time for a single message is $T_{\rm run} \approx 117~\mu$s. Even though $T_{\rm run} \simeq T_1^\text{av}$, the CE task performance using \nisqrc{} on \textit{ibm\_algiers} is essentially independent of qubit lifetimes. This is emphatically demonstrated by the excellent agreement between the experimental results and simulations assuming infinite coherence-time qubits. In fact, finite qubit decay consistent with \textit{ibm\_algiers} leaves simulation results practically unchanged (as plotted in dashed blue); we find that $T_1$ times would have to be over an order of magnitude shorter to begin to detrimentally impact \nisqrc{} performance on this device (see Supplementary Note 7). We further find that artificially increasing $T_{\rm run}$ beyond $T_1$ by introducing controlled delays in each layer also leaves performance unchanged (see Supplementary Note 8).

Using the same device we are able to analyze several important aspects of the \nisqrc{} algorithm. First, we consider the same CE task with a split chain, where the connection between the qubits labelled `14' and `16' on \textit{ibm\_algiers} is severed by removing the $R_{zz}$ gate highlighted in brown in Fig.~\ref{fig:CE_expt_error}(a). The resulting device performance using these two smaller chains is worse, consistent both with simulations of the same circuit and the analogous split Hamiltonian ansatz studied in Results' subsection ``Practical machine learning using temporal data''. Next we return to the $7$~qubit chain but now remove reset operations in the \nisqrc{} architecture, shaded in red in Fig.~\ref{fig:CE_expt_error}(a): all other gates and readout operations are unchanged. The device performance now approaches that of random guessing: the absence of the crucial reset operation leads to an amnesiac QRC with no dependence on past or present inputs. This remarkable finding reinforces that reset operations demanded by the \nisqrc{} algorithm are therefore essential to imbue the QRC with memory and enable any non-trivial temporal data processing. 

We note that for these experiments, while performance qualitatively agrees well with simulations, some quantitative discrepancies are observed. Our deployment of mid-circuit measurements in their earliest implementation on IBM Quantum were accompanied by some technical constraints; for example, not all shots for a given instance of the CE task could be collected in contiguous repeated device runs, instead sometimes being separated by several hours (due to queuing times as well as classical memory buffer constraints on the number of shots that could be collected in a single experiment). Simply put, this means that the device could suffer non-trivial parameter drifts from one type of device configuration to the next, and even during the course of collecting all shots for a specific configuration. In particular, we find that qubit lifetimes for experiments with the split chain, and the connected chain without reset, were significantly shorter than for the connected chain with reset (see Supplementary Tables 1-3), which could lead to the discrepancy in comparison to simulations, where we assumed a fixed coherence time distribution. Resource constraints similarly restrict us to limited training and testing set sizes, which can also lead to variance in performance. We anticipate such technical constraints to be alleviated as mid-circuit measurement implementations mature on IBM Quantum, enabling even more accurate correspondence with simulations.

We also note that there is room for improvement in CE performance when compared against Hamiltonian ansatz \nisqrc{} of similar scale in Fig.~\ref{fig:CE_test_error}. A key difference is the reduced number of connections in the nearest-neighbour linear chain employed on \textit{ibm\_algiers}; including effective $\mathcal{R}_{zz}$ gates between non-adjacent qubits significantly increases the gate-depth of the encoding step, enhancing sensitivity to circuit-fidelity due to increasing runtimes. The circuit ansatz can also be optimized - using knowledge of the Volterra kernels - for better nonlinear processing capabilities demanded by the CE task, in addition to memory capacity determined by $n_{\rm M}$. Nonetheless, the demonstrated performance and robustness of the \nisqrc{} framework to dissipation already suggests its viability for increasingly complex time-dependent learning tasks using actual quantum hardware. 


\section{Discussion}
\label{sec:discussion}

A key technical advancement in our work is the formulation of the Quantum Volterra Theory (QVT) to describe the time-invariant input-output map of a quantum system under temporal inputs and repeated measurements. Insights provided by the QVT enabled us to propose the essential component of the NISQRC algorithm - deterministic post-measurement reset to avoid thermalization due to repeated measurements - which allows the quantum system to retain persistent memory of temporal inputs even under projective measurements and their associated strong backaction. The resulting algorithm enables inference on a signal that can be arbitrarily long, provided the encoding is designed to endow the reservoir with a memory that matches the longest correlation time in the data.

While we have applied the QVT to qubit-based circuits, our analysis does not make an explicit assumption on the Hilbert space dimension of the quantum system, and allows for completely general measurements through its formulation in terms of POVMs; as a result, it can be applied to other finite-level quantum systems such as qudits \cite{kalfus_hilbert_2022}, and can be extended to continuous-variable quantum systems~\cite{nokkala_gaussian_2021, garciabeni_scalable_2023}. We therefore believe the QVT provides the ideal framework to analyze the memory and computational capacity of temporal information processing schemes using general quantum systems and their associated measurement protocols. We note here that the use of continuous weak measurements, analyzed in Ref.~\cite{mujal_time-series_2023}, provides an alternative approach to endowing the reservoir with a finite persistent memory and can be analyzed with QVT for its task-specific optimization.

Going beyond the crucial reset component, we have demonstrated that QVT can be invaluable in identifying general design principles for qubit-based systems as reservoirs. For example, while measuring some fraction of qubits is essential for extracting information, measuring all qubits imposes a trivial memory time. We employ $M \simeq R$ in this work, but an optimal separation of memory and readout qubits may depend on specific tasks.
A simple rule of thumb is to choose $M/R$, together with $\tau$ and other drive strengths, to match the memory time of the physical system to the longest correlation time in the data. In the channel equalization task studied in Results' subsection ``Practical machine learning using temporal data'', the data correlation time is fixed by the choice of the distorting channel and we have then chosen $M/R$ to endow the quantum system with a memory time -- calculated through the QVT formalism -- that matches that time, about $8$ steps (recall memory time is measured in number of encoding steps). Especially, the duration of the unmonitored dynamics, $\tau$, has been chosen to be long enough to generate non-linear kernels that match the known order of the non-linearity of the distorting channel, but short enough to avoid limiting the memory time by the shortest $T_1$. For the latter requirement, the kind of analysis shown in Fig.\,\ref{fig:nM_T1}(c), calculated through QVT, can act as a very helpful guide. We have observed that even when there is a large spread in $T_1$, the physical memory time may be longer than the shortest $T_1$, presumably through the delocalization of the information on longer-lived memory qubits in the circuit. 
Additionally, qubit connectivity, analyzed in Fig.\,\ref{fig:CE_test_error}(a), can help with the generation of functions that are sufficiently complex to match the functional complexity of the task.

Finally, the most crucial design criterion for any quantum system intended to process streaming data is that the map $\mathcal{P}_0$ be non-unital. In an architecture with memory and readout qubits, the presence of a reset operation is essential but not on its own sufficient: as noted earlier the quantum channel must additionally contain input-dependent operations on both memory and readout qubits to prevent scrambling of memory qubits and endow the QRC state with the fading memory property.
To address an important example, it is straightforward to confirm that any channel with input-dependent operations on only memory qubits $\mathcal{U}(u_n)$ and an arbitrary set of controlled-gates from memory to readout qubits (e.g.~Fig.\,5(d) in Ref.\,\cite{yasuda_quantum_2023}) is unital on memory qubits and therefore lacks a persistent temporal memory. We note that in such cases the reservoir can still be trained to implement its function in the transient \cite{brunner_parallel_2013}, but genuine online learning will not be possible. The QVT presented here prescribes how to avoid such pitfalls when designing a quantum channel for temporal data processing: one can simply check whether the resulting $\mathcal{P}_0$ is a unital map.
We have not carried out here an exhaustive study of the optimal design principles for more complex or general classes of tasks, but we hope that the simple and fundamental guidelines we have followed for designing an experimental reservoir to accurately carry out equalization on RF-encoded messages illustrates the utility of QVT in the design of a hardware reservoir.


By enabling online learning in the presence of losses, \nisqrc{} paves the way to harness quantum machines for temporal data processing in far more complex applications than the CE task demonstrated here. Examples include spatiotemporal integrators, and ML tasks where spatial information is temporally encoded, such as video processing. Recent results provide evidence that the most compelling applications however lie in the domain of machine learning on weak signals originating from other, potentially complex quantum systems~\cite{angelatos_reservoir_2021, khan_physical_2021} for the purposes of quantum state classification. In tackling such increasingly complex tasks, the scale of quantum devices required is likely to be larger than those employed here. The NISQRC framework can be applied irrespective of device size; however, its readout features at a given time live in a $K=2^R$ dimensional space. For applications requiring a large $R$, the exponential growth of the feature-space dimension may give rise to concerns with under-sampling, as in practice the available number of shots $\NS$ may not be sufficiently large. In such large-$R$ regimes, certain linear combinations of measured features can be found, known as \textit{eigentasks}, that provably maximize the SNR~\cite{hu_tackling_2023} of the functions approximated by a given physical quantum system trained with $\NS$ shots. Eigentask analysis provides very effective strategies for noise mitigation. In Ref.~\cite{hu_tackling_2023} the {\it Eigentask Learning} methodology was proposed to enhance generalization in supervised learning. For the present work, such noise mitigation strategies were not needed as the size of the devices used were sufficiently small to efficiently sample. An interesting direction is the application of Eigentask analysis to NISQRC, which we leave to future work.

The present work, and the availability of an algorithm for information processing beyond the coherence time, presents new opportunities for mid-circuit measurement and control. While mid-circuit measurement is essential for quantum error correction~\cite{acharya_suppressing_2023}, its recent availability on cloud-based quantum computers has allowed exploration of other quantum applications on near-term noisy qubits. 
Local operations such as measurement followed by classical control for gate teleportation have been used to generate nonlocal entanglement~\cite{zhou_methodology_2000, baumer_efficient_2023, bluvstein_logical_2023}. Additionally, mid-circuit measurements have been employed to study critical phenomena such as phase transitions~\cite{haghshenas_probing_2023, chertkov_characterizing_2023, chen_realizing_2023} and are predicted to allow nonlinear subroutines in quantum algorithms~\cite{holmes_connecting_2022}. The present work opens up a new direction in this application space, namely the design of self-adapting circuits for inference on temporal data with slowly-changing statistics. This would require dynamic programming capabilities for mid-circuit measurements, not employed in the present work. We show here that implementing even the relatively simple CE task challenges current capabilities for repeated measurements and control; having a means to deploy more complex quantum processors for temporal learning via \nisqrc{} can push hardware advancements to more tightly integrate quantum and classical processing for efficient machine-based  inference.


\section{Methods}

\subsection{Generating features via conditional evolution and measurement}
\label{app:methodFeature}
Here we detail how an input-output functional map is obtained in the \nisqrc{} framework. The quantum system is initialized to $\hat{\rho}^{\mathsf{MR}}_0 = \hat{\rho}^{\mathsf{M}}_0 \otimes \ket{0} \! \bra{0}^{\otimes R}$, where $\hat{\rho}^{\mathsf{M}}_0$ is the initial state, which is usually set to be $\ket{0} \! \bra{0}^{\otimes M}$. Then, for each run or `shot' indexed by $s$, the process described in the following paragraph is repeated. 

Before executing the $n$-th step, the overall state can be described as $\hat{\rho}^{\mathsf{M,\mathtt{cond}}}_{n-1} \otimes \ket{0}\!\bra{0}^{\otimes R}$ (usually pure), where the superscript $\mathtt{cond}$ emphasizes that the memory subsystem state is generally conditioned on the history of all previous inputs $\{u_m\}_{m\leq n-1}$ and all previous stochastic measurement outcomes. The readout subsystem state is in a specific pure state, which can be ensured by the deterministic reset operation we describe shortly. Then, the current input $u_n$ is encoded in the quantum system via the parameterized quantum channel $\mathcal{U}(u_n)$, generating the state $\hat{\rho}^{\mathsf{MR},\mathtt{cond}}_n = \mathcal{U}(u_n) \big(\hat{\rho}^{\mathsf{M,\mathtt{cond}}}_{n-1} \otimes \ket{0}\!\bra{0}^{\otimes R} \big)$. 
In this work, $\mathcal{U}(u_n)$ takes the form of continuous evolution under Eq.~\eqref{eq:dissipation} for a duration $\tau$, or the discrete gate-sequence $\hat{U}(u_n)$ depicted in Fig.~\ref{fig:CE_expt_error}(a). The $R$ readout qubits are then measured per Eq.~\eqref{eq:nisqrcPOVM}, and the observed outcome is represented as an $R$-bit string: $\mathbf{b}^{(s)}(n) = (b^{(s)}_{M+1}(n), \cdots, b^{(s)}_{M+R}(n) )$.  Here we consider simple `computational basis' (i.e.~$\hat{\sigma}^z$) measurements, where each bit simply denotes the observed qubit state. A given outcome $j$ occurs with conditional probability $\mathrm{Tr} ( \hat{M}_j \hat{\rho}^{\mathsf{MR},\mathtt{cond}}_n)$ as given by the Born rule, and the quantum state collapses to the new state $\hat{\rho}^{\mathsf{M},\mathtt{cond}}_n \otimes \ket{\mathbf{b}_j}\!\bra{\mathbf{b}_j}$ associated with this outcome. Finally, all $R$ readout qubits are deterministically reset to the ground state (regardless of the measurement outcome); the quantum system is therefore in state $\hat{\rho}^{\mathsf{M},\mathtt{cond}}_{n} \otimes \ket{0}\!\bra{0}^{\otimes R}$.  This serves as the initial state into which the next input $u_{n+1}$ is encoded, and the above process is iterated until the entire input sequence $\bf{u}$ is processed. It is important to notice that $\hat{\rho}^{\mathsf{M}}_{n}$ depends on the observed outcome in step $n-1$ and thus the quantum state and its dynamics for a specific shot is conditioned on the history of measurement outcomes $\{b_i^{(s)}(m)\}_{m < n}$.

By repeating the above process for $S$ shots, one obtains what is effectively a histogram of measurement outcomes at each time step $n$ as represented in Fig.~\ref{fig:schematic}.  The output features are taken as the frequency of occurrence of each measurement outcome, as in Ref.~\cite{hu_tackling_2023}: $\bar{X}_j(n) = \frac{1}{S} \sum_{s=1}^{S} X_{j}^{(s)}(n; \UI)$, where $X_{j}^{(s)}(n; \UI) = \delta(\mathbf{b}^{(s)}(n), \mathbf{b}_j)$ counts the occurrence of outcome $j$ at time step $n$.  These features are stochastic unbiased estimators of the underlying quantum state probability amplitudes $x_j(n) = \Es{ X_j^{(s)}(n; \UI)} = \lim_{S \to \infty} \bar{X}_j(n)$ \cite{hu_tackling_2023}.  As noted in the main text, the final \nisqrc{} output is obtained by applying a set of time-independent linear weights to approximate the target functional $\bar{y}_n = \mathbf{w} \cdot \bar{\mathbf{X}} (n)$. Importantly, during each shot $s\in[S]$, we execute a circuit with depth $N$; the total processing time is therefore $O(N S)$.  If instead one re-encoded $N_m$ previous inputs prior to each successive measurement the processing time is $O(N_m N S)$: $N_m=O(N)$ if the entire past sequence is re-encoded as is conventionally done in QRC \cite{fujii_harnessing_2017, chen_temporal_2020, mujal_time-series_2023}.

\subsection{The quantum Volterra theory and analysis of \nisqrc{}}
\label{app:methodsVolterra}

At any given time step $n$, the conditional dependence on previous measurement outcomes, presented in Methods' subsection ``Generating features via conditional evolution and measurement'', is usually referred to as {\it backaction}. Defining $\hat{\rho}^{\mathsf{MR}}_n$ as the effective pre-measurement state of the quantum system at time step $n$ of the \nisqrc{} framework, quantum state evolution from time step $n-1$ to $n$ can be written via the maps:
\begin{align}
    \hat{\rho}_{n}^{\mathsf{MR}} & = \mathcal{U}(u_n) \left( \mathrm{Tr}_{\mathsf{R}} (\hat{\rho}_{n-1}^{\mathsf{MR}}) \otimes \ket{0} \! \bra{0}^{\otimes R} \right), \\
    \hat{\rho}_{n}^{\mathsf{M}} & = \mathrm{Tr}_{\mathsf{R}} ( \mathcal{U}(u_n) ( \hat{\rho}_{n-1}^{\mathsf{M}} \otimes \ket{0} \! \bra{0}^{\otimes R} )) \equiv \mathcal{C}(u_n) \hat{\rho}_{n-1}^{\mathsf{M}},
    \label{eq:one-step-evol}
\end{align}
which describes the reset of the post-measurement readout subsystem after time step $n-1$, followed by input encoding via $\mathcal{U}(u_n)$ into the full quantum system state. With an eye towards the construction of an I/O map, it proves useful to introduce the expansion of the relevant single-step maps $\mathcal{U}(u)$ and $\mathcal{C}(u)$ in the basis of input monomials $u^k$: $\mathcal{U} (u) \hat{\rho}^{\mathsf{MR}} = \sum_{k = 0}^{\infty} u^k \mathcal{R}_k \hat{\rho}^{\mathsf{MR}}$ and $\mathcal{C} (u) \hat{\rho}^{\mathsf{M }} = \sum_{k = 0}^{\infty} u^k \mathcal{P}_k \hat{\rho}^{\mathsf{M }}$. Then, via iterative application of \Eq{eq:one-step-evol}, $\hat{\rho}^{\mathsf{MR}}_n$ can be written as:
\begin{align}
    \hspace{-1mm} \hat{\rho}^{\mathsf{MR}}_n = & \hspace{-4mm} \sum_{k_1, \hspace{-0.5mm}\cdots\hspace{-0.5mm}, k_n = 0}^{\infty} \hspace{-4mm} u_1^{k_1} \!\! \cdots \! u_n^{k_n} \mathcal{R}_{k_n} \hspace{-1mm} \left( \mathcal{P}_{k_{n-1}} \!\! \cdots \! \mathcal{P}_{k_1} \hat{\rho}^{\mathsf{M}}_0 \! \otimes \! \ket{0} \! \bra{0}^{\otimes R} \right)\!. \hspace{-1mm} \label{eq:rho^MR_n}
\end{align}
The measured features $x_j(n)$ can then be obtained via $x_j(n) = \mathrm{Tr} ( \hat{M}_j \hat{\rho}^{\mathsf{MR}}_n)$.

In the Supplementary Note 3, we show that these $x_j(n)$ obtained using the \nisqrc{} framework can indeed be expressed as a Volterra series
\begin{align}
    \hspace{-2mm} x_j (n) = \sum_{k = 0}^{\infty} \sum_{n_1 = 0}^{\infty} \cdots \hspace{-3mm} \sum_{n_k = n_{k-1}}^{\infty} \hspace{-3mm} h_k^{(j)} (n_1, \cdots, n_k) \prod_{\kappa = 1}^{k} u_{n - n_{\kappa}} \label{eq:Volterra_method}
\end{align}
in the infinite-shot limit. The existence of this manifestly time-invariant form is only possible due to the existence of an information steady-state, guaranteed for a quantum mechanical system under measurement.    

Due to fading memory, the Volterra kernel $h_k^{(j)} (n_1, \cdots, n_k)$ characterizes the dependence of the systems' output at time $n$ on inputs at most $n_k$ steps in the past (recall $n_1 \leq \cdots \leq n_k$, see \Eq{eq:Volterra_method}). The evolution of $\hat{\rho}^{\mathsf{MR}}_n$ upto step $n-n_k$, namely for all $i < n - n_k$, is thus determined entirely by the null-input superoperator $\mathcal{P}_0$. Then the existence of a Volterra series simply requires the existence of an asymptotic steady state for the memory subsystem, $\lim_{n \to \infty} \mathcal{P}^{n}_0 \hat{\rho}^{\mathsf{M}}_0 = \hat{\rho}^{\mathsf{M}}_{\mathrm{FP}}$. As shown in the Supplementary Note 3, such a fixed point is usually ensured by the map $\mathcal{P}_0 \hat{\rho}^{\mathsf{M}} = \mathcal{C}(0) \hat{\rho}^{\mathsf{M}} = \mathrm{Tr}_{\mathsf{R}} ( \mathcal{U}(0) (\hat{\rho}^{\mathsf{M}} \otimes \ket{0} \! \bra{0}^{\otimes R}) )$ being a CPTP map in generic quantum systems. This immediately indicates the fundamental importance of $\mathcal{P}_0$, the operator that corresponds to the single-step map of the memory subsystem under null input: it determines the ability of the \nisqrc{} framework to evolve the quantum system to a unique statistical steady state, guaranteeing the asymptotic time-invariance property, and hence the existence of the Volterra series. 

One byproduct of computing infinite-$S$ features $\{ x_j (n) \}$ is that it enables us to approximately simulate $\{ \bar{X}_j (n) \}$ in a very deep $N$-layer circuit for finite $S$, without sampling individual quantum trajectories under $N$ repeated projective measurement described in Methods' subsection ``Generating features via conditional evolution and measurement''. In fact, given any $n$, once we evaluate a probability distribution $\{ x_j (n) \geq 0 \}$ satisfying $\sum_j x_j (n) = 1$, we can i.i.d.\,sample under this distribution vector for $S$ shots and construct the frequency $\{ \tilde{X}_j (n) \}$ as an approximation of $\{ \bar{X}_j (n) \}$. The validity of this approximation is ensured by the additive nature of loss functions in the time dimension. More specifically, given $Q$ input sequences $\{ \mathbf{u}^{(q)} \in [-1, 1]^N \}_{q \in [Q]}$, a general form of loss function is $\mathscr{L} = \frac{1}{Q N} \sum_q \sum_n \mathcal{L} (\bar{\mathbf{X}} (n ; \mathbf{u}^{(q)}))$. As shown in Appendix C5 of Ref.\,\cite{hu_tackling_2023}, $\frac{1}{Q} \sum_q \mathcal{L} (\bar{\mathbf{X}} (n ; \mathbf{u}^{(q)})) \approx \frac{1}{Q} \sum_q \mathcal{L} (\tilde{\mathbf{X}} (n ; \mathbf{u}^{(q)}))$ in all orders of $\frac{1}{S}$-expansion for any $n \in [N]$, as long as $Q$ is large enough. This is because the probability distribution of $\{ \tilde{X}_j (n) \}$ is exactly the same as the distribution (marginal in time slice) of $\{ \bar{X}_j (n) \}$. Therefore, $\frac{1}{Q N} \sum_q \sum_n \mathcal{L} (\tilde{\mathbf{X}} (n ; \mathbf{u}^{(q)}))$ is a good approximation of $\mathscr{L}$.

In Supplementary Note 2 and Supplementary Note 3, we show that without the reset operation, the fixed-point memory subsystem density matrix is the identity, $\hat{\rho}^{\mathsf{MR}}_{\mathrm{FP}} = \hat{I}^{\otimes L}/2^L$. While this steady state is independent of the initial state and therefore possesses a fading memory, it can be shown that the I/O map it enables is entirely independent of all past inputs as well, so that Volterra kernels $h_k^{(j)} = 0$ for any $k \leq 1$. This yields a trivial reservoir, unable to provide any response to its inputs $u$. Such single-step maps $\mathcal{C}(u)$ are referred to as \textit{unital maps} (maps that map identity to identity), and must be avoided for the \nisqrc{} architecture to approximate any nontrivial functional. The inclusion of reset serves this purpose handily, although we have found certain improper encodings with reset to still result in unital maps $\mathcal{C}(u)$ (e.g., setting $n_T=1$ in the circuit ansatz depicted in Fig.\,\ref{fig:CE_expt_error}(a)). 

A more rigorous sufficient condition for obtaining a nontrivial functional map, referred to as \textit{fixed-point non-preserving} map in the main text, is that $\mathcal{C}(u)$ does not share the same fixed points for all $u$. It is equivalently $\mathcal{P}_k \hat{\varrho}^{\mathsf{M}}_{\mathrm{FP}} \neq 0$ for some $k\geq 1$, due to the identity $\mathcal{C}(u) \hat{\varrho}^{\mathsf{M}}_{\mathrm{FP}} = \hat{\varrho}^{\mathsf{M}}_{\mathrm{FP}} + \sum_{k=1}^{\infty} u^k \mathcal{P}_k \hat{\varrho}^{\mathsf{M}}_{\mathrm{FP}}$. We will prove the importance of this criteria in Supplementary Note 3. The breaking of this criteria will lead to a memoryless reservoir for all earlier input steps: if $\mathcal{P}_k  \hat{\rho}^{\mathsf{M}}_{\mathrm{FP}} = 0$ for all $k \geq 1$, then $h^{(j)}_k(n_1, n_2, \cdots, n_k) \neq 0$ only if $n_1=n_2=\cdots=n_k=0$. A similar result for quantum reservoirs characterized by quantum channels can also be found from Theorem 2 in Ref.\,\cite{martinez-pena_quantum_2023}.

\subsection{Spectral theory of \nisqrc{}: memory, measurement, and kernel structures}
\label{subsec:spectral}

Recall that we can always define the spectral problem $\mathcal{P}_0 \hat{\varrho}^{\mathsf{M}}_\alpha = \lambda_\alpha \hat{\varrho}^{\mathsf{M}}_\alpha$ where $\hat{\varrho}^{\mathsf{M}}_\alpha$ are eigenvectors that exist in the $(2^M)^2 = 4^M$-dimensional space of memory subsystem states, and whose eigenvalues satisfy $1 = \lambda_1 \geq |\lambda_2| \geq \cdots \geq |\lambda_{4^M}| \geq 0$. The importance of the spectrum of $\mathcal{P}_0$ is obvious from the definition of $\hat{\rho}^{\mathsf{M}}_{\mathrm{FP}}$ already. As $\hat{\rho}^{\mathsf{M}}_{\mathrm{FP}}$ is the fixed point of the map defined by $\mathcal{P}_0$, it must equal the eigenvector $\hat{\varrho}^{\mathsf{M}}_1$ since $\lambda_1=1$. Then writing the initial density matrix in terms of these eigenvectors, $\hat{\rho}^{\mathsf{M}}_0 = \sum_{\alpha} d_{0\alpha} \hat{\varrho}^{\mathsf{M}}_\alpha$, the fixed point becomes $\hat{\rho}^{\mathsf{M}}_{\mathrm{FP}} =  \lim_{n\to\infty} \left( \hat{\varrho}^{\mathsf{M}}_1 + \sum_{\alpha\geq 2} d^0_{\alpha} \lambda_{\alpha}^n \hat{\varrho}^{\mathsf{M}}_\alpha \right)$.
This not only reproduces the result $\lim_{n \to \infty} \mathcal{P}^{n}_0 \hat{\rho}^{\mathsf{M}}_0 = \hat{\rho}^{\mathsf{M}}_{\mathrm{FP}}$ but also shows that the approach to the fixed point $\hat{\rho}^{\mathsf{M}}_{\mathrm{FP}} = \hat{\varrho}^{\mathsf{M}}_1$ must be determined by the magnitude of $\lambda_2$; the smaller the magnitude, the faster terms for $\alpha \geq 2$ decay and hence the shorter the memory time. 

To see more directly how the spectrum of $\mathcal{P}_0$ influences memory of inputs, it is sufficient to analyze the Volterra kernels in \Eq{eq:Volterra}. Focusing on single-time contributions from $u_{n-p}$ to $x_j(n)$ at all orders of nonlinearity (multi-time contributions are exponentially suppressed, see Supplementary Note 4), these may be expressed as 
\begin{align}
    \sum_{k=1}^{\infty} \! h_k^{(j)}\!(p^{\otimes k}) \, u^k_{n - p} = \sum_{\alpha = 2}^{4^M} \nu^{(j)}_{\alpha} \lambda^{p - 1}_\alpha F_\alpha(u_{n-p}), \label{eq:internal_features}
\end{align}
which can be viewed as a spectral representation of Volterra kernel contributions to the $j$th measured feature obtained via POVM $\hat{M}_j$. Here, $F_\alpha (u) = \sum_{k = 1}^{\infty} c^{(k)}_{\alpha 1} u^k$ define $4^M-1$ \textit{internal features}, so-called as they depend only on input encoding operators via $\mathcal{P}_k \hat{\varrho}^{\mathsf{M}}_{\alpha'} = \sum_{\alpha = 2}^{4^M} c_{\alpha \alpha'}^{(k)} \hat{\varrho}^{\mathsf{M}}_{\alpha}$, and are in particular independent of the measurement scheme.
Nontrivial $F_\alpha (u)$ and $c_{\alpha 1}^{(k)}$ can be guaranteed if $\mathcal{P}_k \hat{\varrho}^{\mathsf{M}}_{\mathrm{FP}} \neq 0$ for some $k\geq 1$.
The dependence of observables on the measurement basis is via coefficients $\nu^{(j)}_{\alpha} = \mathrm{Tr} ( \hat{M}_j \mathcal{R}_0 ( \hat{\varrho}^{\mathsf{M}}_\alpha \otimes \ket{0} \! \bra{0}^{\otimes R} ) )$. Crucially, the weighting of $F_{\alpha}(u_{n-p})$ for $p$ steps in the past is determined by eigenvalues $\lambda_{\alpha}^{p-1}$ of $\mathcal{P}_0$. For each $\alpha \geq 2$, it vanishes when we take long time limit $p\to\infty$. This property is usually referred as \textit{fading memory}. It also clearly defines a set of distinct, but calculable, memory fading rates $\{|\lambda_{\alpha}|\}_{\alpha \geq 2}$.

Importantly, the ability to construct Volterra kernels and internal features enable us to approximately treat the infinite-dimensional function $x_j(n) = \mathcal{F}_j(u_{\leq n})$ as a function with support only over a space with \textit{effective task dimension} $d_{\mathrm{eff}} = O(n_{\rm M})$, representing $d_{\mathrm{eff}}$ time steps in the past:
\begin{equation}
    x_j(n) = \mathcal{F}_j(u_{\leq n}) \approx \mathcal{F}_j(u_{n - d_{\mathrm{eff}}}, \cdots, u_{n-1}, u_{n}),
\end{equation}
and we can interpret the fading memory functional as a function: $y(n) \approx \mathcal{F}(u_{n-d_{\mathrm{eff}}}, \cdots, u_{n-1}, u_{n})$. In other words, at any given time \nisqrc{} can approximate nonlinear functions that live in a domain of dimension $d_{\mathrm{eff}}$.

\subsection{IBM Quantum implementation}
\label{app:methodsibm}

We recall that the encoding circuit $\hat{U}(u_n) = \left(\mathcal{W}(J)  \mathcal{R}_{z}(\bm{\theta}^z +\bm{\theta}^I u_n ) \mathcal{R}_{x}(\bm{\theta}^x) \right)^{n_{T}}$ for the experimental IBM Quantum implementation in Results' subsection ``Experimental results on quantum system'' describes a composite set of single and two-qubit gates repeated $n_T$ times. Here $\mathcal{R}_{x,z}$ are composite Pauli-rotations applied qubit-wise, e.g.~$ \mathcal{R}_{z} = \bigotimes_{i} \hat{R}_{z}({\theta}^z_i +{\theta}^I_i u) $. $\mathcal{W}(J)$ defines composite two-qubit coupling gates, $\mathcal{W}(J) = \prod_{\langle i, i' \rangle} \mathcal{W}_{i, i'}(J) = \prod_{\langle i, i' \rangle} \mathrm{exp}\{- i (J\tau/n_T) \hat{\sigma}^z_{i} \hat{\sigma}^z_{i'}\} $ for neighboring qubits $i$ and $i'$ along a linear chain in the device and some fixed $J$. The rotation angles $\theta^{x,z,I}$ are randomly drawn from a positive uniform distribution with limits $[a,a+\delta]$, where $a = \frac{\tau}{n_{T}}\theta_{\mathrm{min}}^{x,z,I}$  and $\delta = \frac{\tau}{n_{T}}\Delta \theta^{x,z,I}$. We find that letting the number of Trotterization steps $n_{T}=3$ is sufficient to generate a well-behaved null-input CPTP map $\mathcal{P}_0$. Our hyperparameter choices are further tuned to ensure a memory time $n_{\rm M}$ commensurate with the CE task dimension. The particular hyperparameter choices for the plot in Fig.~\ref{fig:CE_expt_error} are $\theta^{x,z,I}_{\rm min}= \{1.0,0.5,0.1\}$, $\Delta\theta^{x,z,I} = \theta^{x,z,I}_{\rm min}$, $J=1$, $n_{T} = 3$, and $\tau = 1$. 

In the experiment, mid-circuit measurements and qubit resets are performed as separate operations, due to the differences in control flow paths between returning a result and the following qubit manipulation~\cite{hua_exploiting_2022}. Related hardware complexities restrict us to a slightly shorter instance of the CE task than considered in Results' subsection ``Practical machine learning using temporal data'', with messages $m(n)$ of length $N=20$, submitted in batches of 200 jobs with 100 circuits each and 125 observations (shots) per circuit in order to prevent memory buffer overflows. Regardless, using cross-validation techniques, we ensure that our observed training and testing performance is not influenced by limitations of dataset size. We also forego the initial washout period needed to reach $\rho^{\mathsf{MR}}_{\rm FP}$ for similar reasons. Finally, the $\mathcal{W}_{i, i'}(J)$ rotations in the two-qubit Hilbert space that implement $\mathcal{W}(J)$ are generated by the native echoed cross-resonance interaction of IBM backends~\cite{sheldon_characterizing_2016}, which provides higher fidelity than a digital decomposition in terms of CNOTs for Trotterized circuits~\cite{stenger_simulating_2021}.

\section*{Data Availability}

The data generated for numerical results in this study have been deposited in the Github repository under accession link \url{https://github.com/skhanCC/NISQRC-Codes} \cite{NISQRC-Codes}. The raw experimental data obtained from \textit{ibm\_algiers} are not available in the Github repository due to its huge size, and its access can be be made available to interested parties upon request. The processed experimental data are available at the Github repository. The data of experimental parameters in this study are provided in the Supplementary Note 9. No external data was used in this study.

\section*{Code Availability}

The code used in this article is available in the Github repository \url{https://github.com/skhanCC/NISQRC-Codes}.

\bibliography{bibfile}

\section*{Acknowledgement}

The authors acknowledge the support from the
ARO contract W911NF-19-C0092 (received by G.J.R.),
DARPA contract HR00112190072 (received by H.E.T.), AFOSR award FA9550-20-1-0177 (received by H.E.T.), and AFOSR MURI award FA9550-22-1-0203 (received by H.E.T.). The views, opinions, and findings expressed are solely the authors' and not the U.S. government's. The authors acknowledge the use of IBM Quantum services for this work.


\section*{Author Contributions}

F.H., S.A.K., G.A., and H.E.T.~conceived the project. F.H.~developed the theoretical model, and F.H.~and G.A. and S.A.K performed the numerical simulations. G.E.R.~and G.J.R.~proposed the applications to channel equalization task and the circuit ansatz. S.A.K.~and N.T.B.~performed the experiments on IBM Quantum platform and analyzed the data. H.E.T supervised the project. F.H., S.A.K., G.A.~and H.E.T.~wrote the manuscript based on the contributions from all authors.

\section*{Competing Interests}
The authors declare no competing interests.

\clearpage

\appendix


\begin{widetext}
    \startcontents[Supplementary Information]
{
  \hypersetup{linkcolor=black}
  \printcontents[Supplementary Information]{l}{1}{\section*{Supplementary Information}\setcounter{tocdepth}{2}}
}

\setcounter{page}{1}

\renewcommand{\appendixname}{Supplementary Note}
\renewcommand{\thesection}{\arabic{section}}

\renewcommand{\figurename}{Supplementary Figure}
\setcounter{figure}{0}

\renewcommand{\theequation}{\arabic{equation}}
\setcounter{equation}{0}
\counterwithout{equation}{section}

\renewcommand{\tablename}{Supplementary Table}
\renewcommand{\thetable}{\arabic{table}}

\makeatletter
\let\toc@pre\relax
\let\toc@post\relax
\makeatother


\section{Details of the \nisqrc{} architecture}
\label{sec:NISQRC_Architecture_Detail}


\begin{figure*}[b]
    \centering
    \includegraphics[width=\textwidth]{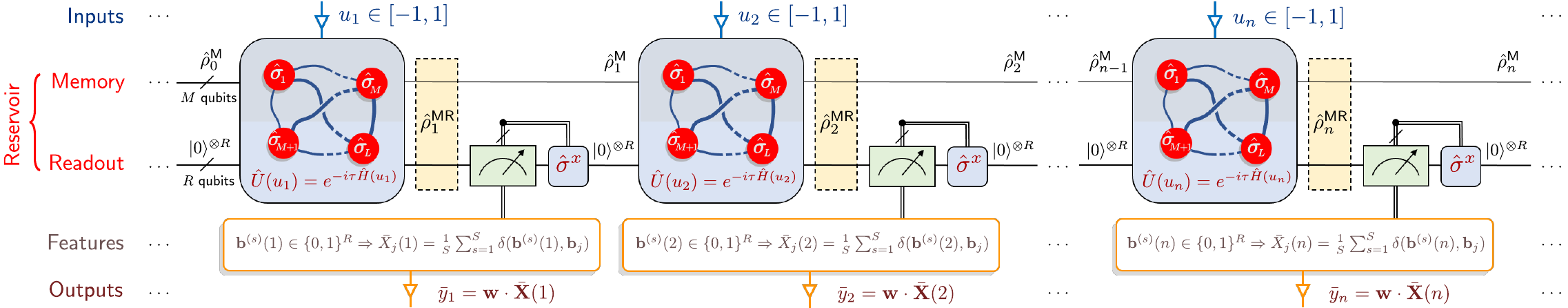}
    \caption{\nisqrc{} architecture to generate a functional map $\mathcal{F}: \UI \mapsto \YO$ by using a qubit-based quantum system. The input function can be written as a time-discrete sequence $\UI = \{u_{-\infty}, \cdots, u_{-1}, u_{0}, u_{1}, \cdots, u_{\infty} \}$, which is encoded in the quantum system at every time step $n$ via a fixed encoding scheme, here shown as a Hamiltonian encoding $\hat{H}(u_n)$. Measured features $\bar{X}_j(n)$ are constructed from finite samples $\NS$ under a specified measurement scheme at each time step (for example probabilities of measured bit-strings $\mathbf{b}^{(s)}(n)$ under computational basis measurement). The output function $\YO = \{y_{-\infty}, \cdots, y_{-1}, y_{0}, y_{1}, \cdots, y_{\infty} \}$ is constructed from these finitely-sampled measured features. The goal of the trained functional $\mathcal{F}$ is to approximate a desired functional $\mathcal{F}^{\star}: \UI \mapsto \YS$, where $\YS = \{y^{\star}_{-\infty}, \cdots, y^{\star}_{-1}, y^{\star}_{0}, y^{\star}_{1}, \cdots, y^{\star}_{\infty} \}$, so that under the same input $\UI$, $y^{\star}_n \approx y_n~\forall~n$ with as little error as possible.}
    \label{fig:QRC_algorithm}
\end{figure*}


\subsection{The \nisqrc{} algorithm}
The underlying dynamical system we analyze in this article consists of $L=M+R$ qubits, with $M$ qubits serving as memory qubits and $R$ qubits serving as readout qubits.
The evolution is governed by a Hamiltonian that is linearly parameterized by a one-dimensional variable $u \in [-1, 1]$ (serving as input): 
\begin{align}
    \hat{H}(u) = \hat{H}_0 + u \cdot \hat{H}_1. \label{eq:Hu=H0+uH1}
\end{align}
We choose a form of $\hat{H}_0$ and $\hat{H}_1$ that can be implemented in a quantum annealing system or analog quantum simulator in a hardware-efficient way: $\hat{H}_0 = \sum_{\langle i, i'\rangle} J_{i,i'} \hat{\sigma}^z_i \hat{\sigma}^z_{i'} + \sum^L_{i=1} \eta^x_{i} \hat{\sigma}^x_i$ and $\hat{H}_1 = \sum^L_{i=1} \eta^z_{i} \hat{\sigma}^z_i$
The coupling strength $J_{i,i'}$, transverse $x$-field strength $\eta^x_{i} = \eta^x + \varepsilon^x_{i}$ and longitudinal $z$-drive strength $\eta^z_{i} = \eta^z + \varepsilon^z_{i}$ are pre-selected via randomness: $J_{i,i'} \sim \mathrm{Unif}[0, J_{\mathrm{max}}]$, $\varepsilon^x_{i} \sim \varepsilon^x_{\mathrm{rms}} \times \mathcal{N}(0, 1)$ and $\varepsilon^z_{i} \sim \varepsilon^z_{\mathrm{rms}} \times \mathcal{N}(0, 1)$. One thing that needs to be emphasized is that the encoding scheme Supplementary Equation \ref{eq:Hu=H0+uH1} is general enough such that encoding Eq.\,(\ref{eq:H0_H1}) is merely an illustrative example. A variety of $\hat{H}_0, \hat{H}_1$ can be employed as long as they are resource-efficiently realized in a physical platform. 

In theory, the domain $\mathbb{Z}$ of $n$ is infinite. However in practical experiments, it is impossible to feed an input sequence from infinite past $n=-\infty$ to infinite future $n=\infty$. Thus we cutoff infinity of time-step index into $n \in [N] \equiv \{1, 2, \cdots, N\}$.
As a summary, now we have a sequence of reservoir recurrent units, each of which is characterized by an underlying Hamiltonian $H(u_n)$ for all $n \in [N]$, and step evolution duration $\tau$. 

As what we will prove in Supplementary Note \ref{app:meas_dynamics},
since calculating readout feature functions $\{ x_j(n) \}_{j \in [K]}$ can be done by taking $x_j(n) = \mathrm{Tr} \! \left( \hat{M}_j \hat{\rho}^{\mathsf{MR}}_n \right)$ where the effective density matrix is $\hat{\rho}^{\mathsf{MR}}_n = \mathcal{U} (u_n) \left( \left( \mathcal{C} (u_{n - 1}) \cdots \mathcal{C} (u_1) \hat{\rho}^{\mathsf{M}}_0 \right) \otimes \ket{0} \! \bra{0}^{\otimes R} \right)$, the full dynamics of \nisqrc{} can also be written into set of recurrent equations
\begin{equation}
\left\{\begin{array}{ll}
    \hspace{1.0mm} \hat{\rho}^{\mathsf{MR}}_n = \, \mathcal{U}(u_n) \! \left( \hat{\rho}^{\mathsf{M}}_{n - 1} \otimes \ket{0}\!\bra{0}^{\otimes R} \right), \vspace{2mm}\\
    \mathbf{x} (n) = \{ x_j(n) \}_{j\in [K]} = \{  \mathrm{Tr} (\hat{M}_j \hat{\rho}^{\mathsf{MR}}_{n} )  \}_{j\in [K]}, \vspace{3mm}\\
    \hspace{3.5mm} y_n = \mathbf{w} \cdot \mathbf{x} (n). 
\end{array}\right. \label{eqs:QRC_arc}
\end{equation}
This algorithm induces a functional $\mathcal{F}: \UI \mapsto \YO$, where $\YO(n) = y_n$. We define an observable
\begin{align}
    \hat{M}_{\mathbf{w}} \equiv \sum^{K-1}_{j=0} w_j \hat{M}_j,
\end{align}
and therefore $y_n = \mathrm{Tr} (\hat{M}_{\mathbf{w}} \hat{\rho}^{\mathsf{MR}}_{n} )$ which affords a great deal of convenience in our notation.

The readout features $x_j(n)$ are nothing but the respective probabilities of measuring $\mathbf{b}_j$ at the $n$-th time step, and we call this readout scheme the \textit{probability representation}~[1]. 
In the literature, the readout features are alternatively chosen to be the quantum spin moments. In this \textit{moment representation}, $\mathcal{O}'_{R} = \{ \hat{M}_j | \hat{M}_j = I^{\otimes M} \otimes \bigotimes^{L}_{i=M+1} \hat{\sigma}_i \}$ where each $\hat{\sigma}_i \in \{\hat{I}, \hat{\sigma}^z\}$. These two different representations can be related by a Walsh-Hadamard transformation~[1]. 

\section{Quantum dynamics under \nisqrc{} -- Role of repeated evolution, measurement, and reset}
\label{app:meas_dynamics}

\subsection{Quantum dynamics under measurement without subsequent qubit reset} 
\label{app:meas_dynamics_1}

For simplicity, we first consider a QRC with $M = 1$ memory qubit and $R = 1$ readout qubit (namely $L = M + R = 2$). Furthermore, we consider $\hat{\sigma}^z$ measurement of the readout qubit at each time step $n$ of the framework. The measurement Kraus operators introduced in the main text then take the specific form
\begin{equation}
    \hat{P}_i = \hat{I} \otimes \ket{i}\!\bra{i}.
\end{equation}
The corresponding observable $M$ can be written as $\hat{M} = \sum^1_{i = 0} i \hat{P}_i = \hat{I} \otimes \ket{1} \! \bra{1}$, which measures the probability of the single readout qubit being in excited state. 

The \nisqrc{} framework then involves a continuous pipeline of evolution under a quite arbitrary superoperator $\mathcal{U}(n)$ (not restricted to the linearly parameterized Hamiltonian form we consider in the main text), followed by measurement, repeated until all inputs $\{u_n\}$ have been processed by the QRC. The inclusion of measurement with stochastic outcomes interleaved with evolution steps, as opposed to at the final step, makes our knowledge of the QRC state conditional on the entire measurement history. For example, starting from the initial state $\hat{\rho}_0^{\mathsf{MR}}$ and evolving under $\mathcal{U}_1$ at time step $n=1$, the subsequent measurement yields a measurement outcome $X_n = i_n$, where $i_n \in \{0,1\}$ for a single readout qubit. The post-measurement state $\hat{\rho}_1^{\mathsf{MR}, \mathtt{cond}}$ is then conditioned on the measurement result at time step $n=1$, as indicated by the superscript $\mathtt{cond}$. For an arbitrary time step $n$, this conditioning thus extends to the entire measurement history $\{X_1,X_2,\ldots,X_{n-1}\}$. The entire pipeline can be viewed schematically as below:
\begin{align}
    \hat{\rho}^{\mathsf{MR}}_0 \xrightarrow{\mathcal{U}_1,\hat{P}_{i_1}} &~\hat{\rho}^{\mathsf{MR}, \mathtt{cond}}_1 \xrightarrow{\mathcal{U}_2,\hat{P}_{i_2}}~\hat{\rho}^{\mathsf{MR}, \mathtt{cond}}_2 \cdots \xrightarrow{\mathcal{U}_n,\hat{P}_{i_n}} \hat{\rho}^{\mathsf{MR}, \mathtt{cond}}_n \cdots \nonumber \\
    &\Downarrow~~~~~~~~~~~~~~~~~~~~~~~~~\Downarrow~~~~~~~~~~~~~~~~~~~~~~~~~~~~~\Downarrow \nonumber \\
    &X_{1}~~~~~~~~~~~~~~~~~~~~~~~~X_{2}~~~~~~~~~~~~~~~~~~~~~~~~~~~~~X_{n}
    \label{eq:EM_chain_1}
\end{align}
It is not hard to show that this process is equivalent to the quantum non-demolition scheme proposed in Ref.\,[2].

In practice, we are often interested not in the result of a single shot, but of the ensemble average computed over many shots; in the limit of infinite-sampling, this defines the readout features $x(n)$ computed via ensemble averages over an infinite number of repeated shots of their stochastic conditional counterparts $X_n$:
\begin{align}
    x(n) = \mathbb{E}[X_n].
\end{align}
Computing this expectation using individual measurement shots would be the standard approach in any experimental \nisqrc{} realization, but is prohibitively expensive for this analysis. This is not least because of the dependence of $X_n$ at any time step $n$ on the entire measurement history $\{X_1,X_2,\ldots,X_{n-1}\}$, a complexity that scales very unfavourably with QRC size and the total number of time steps $N$. Instead, we show that the expectation can be efficiently evaluated - crucially, accounting for the conditional dynamics due to interleaved measurements - to yield a simplified expression for the infinitely-sampled readout features in terms of an effective, ensemble-averaged density matrix $\hat{\rho}_n^{\mathsf{MR}}$, namely $x (n) = \mathrm{Tr} (\hat{M} \hat{\rho}^{\mathsf{MR}}_{n} )$.

To proceed, we note that, by mathematical induction, the conditional state with associated measurement record $\{  X_1 = i_1, \cdots, X_{n - 1} = i_{n - 1} \}$ is
\begin{equation}
    \hat{\rho}^{\mathsf{MR}, \mathtt{cond}}_n = \frac{\mathcal{U}_n \! \left( \hat{P}_{i_{n - 1}} \cdots \mathcal{U}_2 \! \left( \hat{P}_{i_1} \! \left( \mathcal{U}_1 \hat{\rho}^{\mathsf{MR}}_0 \right) \hat{P}^{\dagger}_{i_1} \right) \cdots \hat{P}^{\dagger}_{i_{n - 1}} \right) }{\mathrm{Tr} \left( \hat{P}_{i_{n - 1}} \cdots \mathcal{U}_2 \! \left( \hat{P}_{i_1} \! \left( \mathcal{U}_1 \hat{\rho}^{\mathsf{MR}}_0 \right) \hat{P}^{\dagger}_{i_1} \right) \cdots \hat{P}^{\dagger}_{i_{n - 1}} \right)},
\end{equation}
while the probability of obtaining this measurement record is simply
\begin{align}
    \mathrm{Pr} [ X_1 = i_1, \cdots, X_{n - 1} = i_{n - 1}] = \mathrm{Tr} \! \left( \hat{P}_{i_{n - 1}} \cdots \mathcal{U}_2 \! \left( \hat{P}_{i_1} \! \left( \mathcal{U}_1 \hat{\rho}^{\mathsf{MR}}_0 \right) \hat{P}^{\dagger}_{i_1} \right) \cdots \hat{P}^{\dagger}_{i_{n - 1}} \right) . \label{eq:jp} 
\end{align}
In order to further simplify this expression, we observe the following identity for any $\hat{A} \in \mathbb{C}^{4 \times 4}$, which can be verified by direct computation
\begin{equation}
    \sum_{i = 0, 1} \hat{P}_i \hat{A} \hat{P}^{\dagger}_i = (\hat{\mathbf{1}} \otimes \hat{I}) \circ \hat{A} \label{eq:lempap}
\end{equation}
where the matrices $\hat{I} =
  \left(\begin{array}{cc}
    1 & 0\\
    0 & 1
  \end{array}\right)$ and $\hat{\mathbf{1}} = \left(\begin{array}{cc}
    1 & 1\\
    1 & 1
  \end{array}\right)$,
and the notation $\circ$ represents the \textit{Hadamard product} (element-wise product): $(\hat{A} \circ \hat{B})_{i j} = A_{i j} B_{i j}$. Supplementary Equation \ref{eq:lempap} enables us to introduce the \textit{measurement-induced decoherence superoperator} $\mathcal{M}$:
\begin{equation}
    \mathcal{M} \hat{\rho}^{\mathsf{MR}} =  (\hat{\mathbf{1}} \otimes \hat{I}) \circ \hat{\rho}^{\mathsf{MR}} .
    \label{apxeq:meas}
\end{equation}
Therefore, according to Supplementary Equation \ref{eq:jp}, the unconditional expectation $\mathbb{E} [X_n] = \! \sum_{i_1, \cdots, i_n} \! i_n \mathrm{Pr} [ X_1 = i_1, \cdots, X_{n} = i_{n}]$ of the random variable $X_n$ can be computed by contraction:
\begin{align}
    x(n) = \mathbb{E} [X_n] = \sum_{i_1, \cdots, i_n} \!\!\! i_n \mathrm{Tr} \! \left( \hat{P}_{i_n} \mathcal{U}_n \! \left( \cdots \mathcal{U}_2 \! \left( \hat{P}_{i_1} \! \left( \mathcal{U}_1 \hat{\rho}^{\mathsf{MR}}_0 \right) \hat{P}^{\dagger}_{i_1} \right) \cdots \right) \hat{P}^{\dagger}_{i_n} \right)
    = \, \mathrm{Tr} \! \left( \hat{M} \! \left( \mathcal{U}_n \mathcal{M}\mathcal{U}_{n-1} \cdots \mathcal{U}_2 \mathcal{M}\mathcal{U}_1 \hat{\rho}^{\mathsf{MR}}_0 \right) \right) . \label{eq:contaction}
\end{align}
where we used Supplementary Equation \ref{apxeq:meas} and $\hat{M} = \sum_{i_n} i_n \hat{P}_{i_n}$. This expression naturally leads to the identification of the term in square brackets as the effective density matrix at time step $n$, $\hat{\rho}^{\mathsf{MR}}_n =\mathcal{U}_n \mathcal{M} \cdots \mathcal{U}_2 \mathcal{M}\mathcal{U}_1 \hat{\rho}^{\mathsf{MR}}_0$, such that computing the trace with respect to this density matrix provides any readout feature at time step $n$ in the infinite sampling limit, $x (n) = \mathrm{Tr} (\hat{M} \hat{\rho}^{\mathsf{MR}}_n )$. The generalization to a QRC with $L = M+R \geq 3$ and input sequence $\{ u_n \}$ is now straightforward: $\mathcal{U}_n$ is replaced with $\mathcal{U}(u_n)$, while the measurement-induced decoherence superoperator $\mathcal{M}$ generalizes to: 
\begin{align}
    \mathcal{M} \hat{\rho}^{\mathsf{MR}} = \left( \hat{\mathbf{1}}^{\otimes M} \otimes \hat{I}^{\otimes R} \right) \circ \hat{\rho}^{\mathsf{MR}}.
\end{align}
With these changes, the effective density matrix at time step $n$ for the \nisqrc{} framework without reset is given by 
\begin{align}
    \hat{\rho}_n^{\mathsf{MR}} = \mathcal{U} (u_n) \mathcal{M}\mathcal{U} (u_{n-1})  \cdots \mathcal{U} (u_2) \mathcal{M} \mathcal{U} (u_1) \hat{\rho}_0^{\mathsf{MR}} . \label{eq:UMrho}
\end{align}
Note that $\hat{\rho}^{\mathsf{MR}}_n$ accounts for both any time-dependent unitary dynamics via $\mathcal{U}_n$, as well as the role of repeated measurements via recurrent applications of $\mathcal{M}$.

\subsubsection{Thermalization induced by repeated measurements without reset}

We need to point out that even if the circuits have similar structures to those used in measurement-induced phase transition [3]: at step $n$ associated with unitary evolution $\mathcal{U}_n$, qubits indexed by a random  subset $\mathcal{I}_n \subseteq [L]$ will be measured. In this scenario, the effective state evolution is similar $\hat{\rho}_n^{\mathsf{MR}} =\mathcal{U}_n \mathcal{M}_{\mathcal{I}_{n - 1}} \cdots \mathcal{U}_2 \mathcal{M}_{\mathcal{I}_1} \mathcal{U}_1 \hat{\rho}_0^{\mathsf{MR}}$, the only difference is that measurement-induced decoherence superoperator $\mathcal{M}_{n}$ now is no longer a time-independent map
\begin{align}
    \mathcal{M}_{\mathcal{I}_n} \hat{\rho}^{\mathsf{MR}}  = \left( \bigotimes_{i = 1}^L \hat{E}_i \right) \circ \hat{\rho}^{\mathsf{MR}}, \quad \hat{E}_i = \left\{\begin{array}{ll}
     \hat{I}, & \text{if } i \in \mathcal{I}_n, \\
     \hat{\mathbf{1}}, & \text{if } i \notin \mathcal{I}_n. 
   \end{array}\right.
\end{align}
For any overall state $\hat{\rho}^{\mathsf{MR}}$, the Frobenius distance $\left\| \hat{\rho}^{\mathsf{MR}} - \frac{I^{\otimes L}}{2^L} \right\|^2_F$ will never increase after either unitary evolution $\mathcal{U}$ or measurement $\mathcal{M}_{\mathcal{I}}$:
\begin{align}
    \left\| \mathcal{U} \hat{\rho}^{\mathsf{MR}} - \frac{\hat{I}^{\otimes L}}{2^L} \right\|^2_F & \!\! = \left\| \mathcal{U} \! \left( \hat{\rho}^{\mathsf{MR}} - \frac{\hat{I}^{\otimes L}}{2^L} \right) \right\|^2_F \!\! = \left\| \hat{\rho}^{\mathsf{MR}} - \frac{\hat{I}^{\otimes L}}{2^L} \right\|^2_F, \\
    \left\| \mathcal{M}_{\mathcal{I}} \hat{\rho}^{\mathsf{MR}} - \frac{\hat{I}^{\otimes L}}{2^L} \right\|^2_F & \!\! = \left\| \mathcal{M}_{\mathcal{I}} \! \left( \hat{\rho}^{\mathsf{MR}} - \frac{\hat{I}^{\otimes L}}{2^L} \right) \right\|^2_F \!\! \leq \left\| \hat{\rho}^{\mathsf{MR}} - \frac{\hat{I}^{\otimes L}}{2^L} \right\|^2_F , 
\end{align}
where the proof employs that fully mixed state $\frac{\hat{I}^{\otimes L}}{2^L}$ is the simultaneous fixed point of $\mathcal{U}$ and $\mathcal{M}_{\mathcal{I}}$ (equivalently, both maps are \textit{unital} CPTP map). The non-increasing purity implies that
\begin{align}
    \lim_{n \rightarrow \infty} \hat{\rho}_n^{\mathsf{MR}} = \frac{\hat{I}^{\otimes L}}{2^L}. \label{eq:IL2L}
\end{align}
The final QRC state therefore has no memory of the initial state $\hat{\rho}_0^{\mathsf{MR}}$. 
As a result, in previous works~[4] this type of evolution has been employed to equip QRCs with the fading memory property. However, note that the final state is also entirely independent of the input $u(n)$, which renders it incapable of performing any useful computations on this input.
Hence input-dependent unitary evolution combined with readout only does not yield a useful QRC. We show next how a simple modification of the measurement protocol can allow fading memory without yielding a trivial I/O map.

\begin{figure}[t]
    \centering
    \includegraphics[width = 0.8\columnwidth]{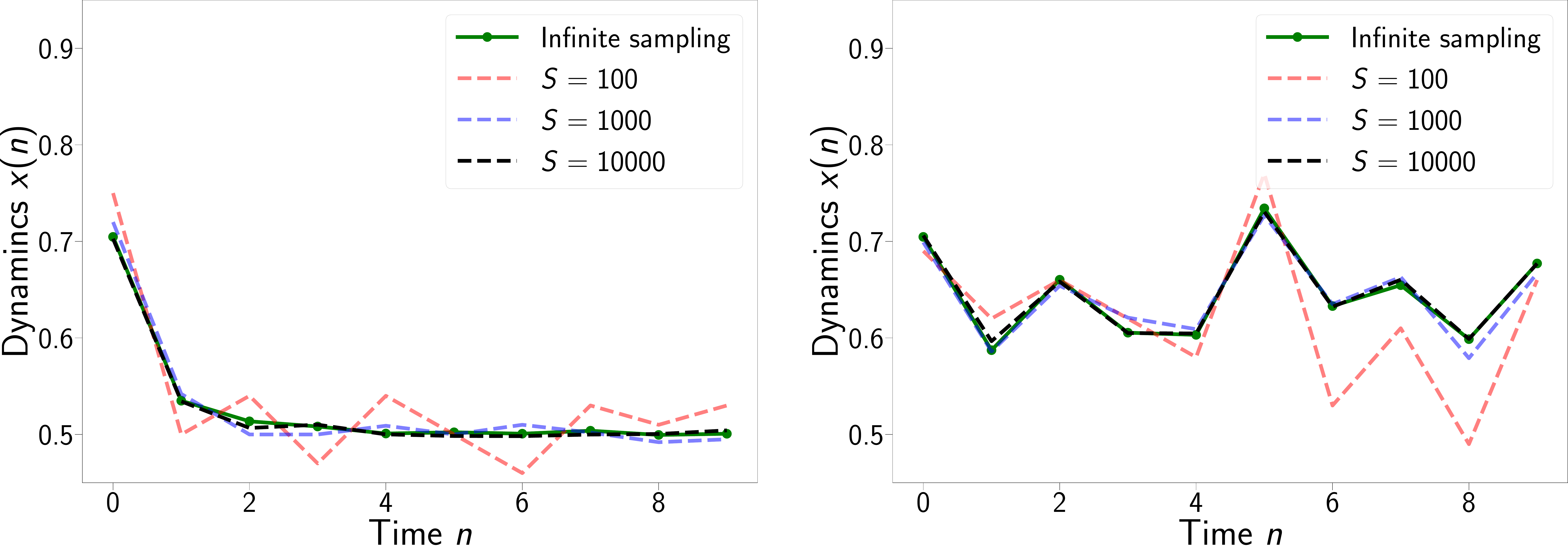}
    \caption{\nisqrc{} readout features for a $(2+1)$-qubit QRC, under both finite sampling (dashed lines) and infinite sampling (solid line and dots). The hyperparameters are $J_{\mathrm{max}} = \eta^x = \varepsilon^x_{\mathrm{rms}} = \eta^z = \varepsilon^z_{\mathrm{rms}} = 1$ in units of $1/\tau$. 
    (Left) Without reset. (Right) With reset. In both cases, with increasing shots, the finitely-sampled readout features become closer to the black dashed features under infinite shots, as expected. However, without reset the readout features approach trivial values dictated by the effective density matrix of Supplementary Equation \ref{eq:IL2L} as $n$ increases.}
    \label{fig:Finite_vs_Infinite_Dynamics_NoReset}
\end{figure}

\subsection{Quantum dynamics under measurement and reset}
\label{app:reset}
For notational simplicity, we once again analyze a system with $M = 1$ memory qubit and $R = 1$ readout qubit (namely $L = M + R = 2$). We apply Pauli $z$ measurement on the readout qubit at each QRC step, the corresponding observable is $\hat{M} = \hat{I} \otimes \ket{1} \! \bra{1}$. Since now we apply the conditional reset. The measurement process is described by a POVM measurement ($i=0,1$): 
\begin{equation}
    \hat{K}_i = \hat{I} \otimes \ket{0} \! \bra{i},
\end{equation}
and thus when overall state $\hat{\rho}^{\mathsf{MR}}$ is measured, the post-measurement state should be $\hat{K}_i \hat{\rho}^{\mathsf{MR}} \hat{K}^{\dagger}_i$ if the random readout index is $i$. These two POVMs satisfy the completeness relation: 
\begin{equation}
    \sum_{i = 0, 1} \hat{K}^{\dagger}_i \hat{K}_i = \sum_{i = 0, 1} \hat{I} \otimes \ket{i} \bra{i} = \hat{I} \otimes \hat{I}.
\end{equation}

The \nisqrc{} pipeline including reset can now be viewed schematically as:
\begin{align}
    \hat{\rho}^{\mathsf{MR}}_0 \xrightarrow{\mathcal{U}_1,\hat{K}_{i_1}} &~\hat{\rho}^{\mathsf{MR}, \mathtt{cond}}_1 \xrightarrow{\mathcal{U}_2,\hat{K}_{i_2}}~\hat{\rho}^{\mathsf{MR}, \mathtt{cond}}_2 \cdots \xrightarrow{\mathcal{U}_n,\hat{K}_{i_n}} \hat{\rho}^{\mathsf{MR}, \mathtt{cond}}_n \cdots \nonumber \\
    &\Downarrow~~~~~~~~~~~~~~~~~~~~~~~~~\Downarrow~~~~~~~~~~~~~~~~~~~~~~~~~~~~~~\Downarrow \nonumber \\
    &X_{1}~~~~~~~~~~~~~~~~~~~~~~~~X_{2}~~~~~~~~~~~~~~~~~~~~~~~~~~~~~~X_{n}
    \label{eq:EM_chain_2}
\end{align}
Proceeding as before, the conditional state with associated measurement record $\{ X_1 = i_1, \cdots, X_{n - 1} = i_{n - 1} \}$ is
\begin{equation}
    \hat{\rho}^{\mathsf{MR}, \mathtt{cond}}_n = \frac{\mathcal{U}_n \! \left( \hat{K}_{i_{n - 1}} \cdots \mathcal{U}_2 \! \left( \hat{K}_{i_1} \! \left( \mathcal{U}_1 \hat{\rho}^{\mathsf{MR}}_0 \right) \hat{K}^{\dagger}_{i_1} \right) \cdots \hat{K}^{\dagger}_{i_{n - 1}} \right) }{\mathrm{Tr} \left( \hat{K}_{i_{n - 1}} \cdots \mathcal{U}_2 \! \left( \hat{K}_{i_1} \! \left( \mathcal{U}_1 \hat{\rho}^{\mathsf{MR}}_0 \right) \hat{K}^{\dagger}_{i_1} \right) \cdots \hat{K}^{\dagger}_{i_{n - 1}} \right)},
\end{equation}
and the probability of obtaining this measurement record $\{ X_1 = i_1, \cdots, X_{n} = i_{n} \}$ is
\begin{align}
    \mathrm{Pr} [X_1 = i_1, \cdots, X_n = i_n] = \mathrm{Tr} \! \left( \hat{K}_{i_{n}} \mathcal{U}_{n} \! \left( \hat{K}_{i_{n-1}} \mathcal{U}_{n-1} \! \left( \cdots \hat{K}_{i_1} \! \left( \mathcal{U}_1 \hat{\rho}^{\mathsf{MR}}_0 \right) \hat{K}^{\dagger}_{i_1} \cdots \right) \hat{K}_{i_{n-1}} \right) \hat{K}^{\dagger}_{i_{n}} \right) .
\end{align}
which are analogous to the previous results with the replacement $\hat{P}_{i_n} \to \hat{K}_{i_n}$. Similar to Supplementary Equation \ref{eq:lempap}, we can verify that for any $\hat{A} \in \mathbb{C}^{4 \times 4}$,
\begin{align}
    \sum_{i=0,1} \hat{K}_i \hat{A} \hat{K}^{\dagger}_i = \mathrm{Tr}_{\mathsf{R}} (\hat{A}) \otimes \ket{0}\!\bra{0} . 
\end{align}

For a quantum reservoir, we let $\mathcal{U}_n = \mathcal{U}(u_n)$. A similar contraction as Supplementary Equation \ref{eq:contaction}
\begin{align}
    \sum_i \hat{K}_i \mathcal{U} (\hat{\rho}^{\mathsf{M}} \otimes \ket{0}\!\bra{0}) \hat{K}^{\dagger}_i = \mathrm{Tr}_{\mathsf{R}} (\mathcal{U} (\hat{\rho}^{\mathsf{M}} \otimes \ket{0}\!\bra{0})) \otimes \ket{0}\!\bra{0}
\end{align}
gives the effective state evolution $x (n) = \mathrm{Tr}( \hat{M} \hat{\rho}^{\mathsf{MR}}_n )$ where $\hat{\rho}^{\mathsf{MR}}_n = \mathcal{U}_n \left( \left( \mathcal{C}_{n - 1} \cdots \mathcal{C}_1 \hat{\rho}^{\mathsf{M}}_0 \right) \otimes \ket{0}\!\bra{0} \right)$ and $\mathcal{C}_{n} \rho^{\mathsf{M}} = \mathrm{Tr}_{\mathsf{R}} \! \left( \mathcal{U}_{n} \! \left( \hat{\rho}^{\mathsf{M}} \otimes \ket{0}\!\bra{0} \right) \right)$.

Also, for more general $M>1$ and $R>1$ we used in the main text, we can still introduce the effective density matrices $\hat{\rho}^{\mathsf{MR}}_n$ in \nisqrc{} having the same expression
\begin{align}
    \hat{\rho}^{\mathsf{MR}}_n =\mathcal{U} (u_n) \left( \left( \mathcal{C} (u_{n - 1}) \cdots \mathcal{C} (u_1) \hat{\rho}^{\mathsf{M}}_0 \right) \otimes \ket{0}\!\bra{0}^{\otimes R} \right). \label{apxeq:rhoMRn}
\end{align}
where $\mathcal{C} (u) \hat{\rho}^{\mathsf{M}} = \mathrm{Tr}_{\mathsf{R}} \! \left( \mathcal{U} (u) \left( \hat{\rho}^{\mathsf{M}}_0 \otimes \ket{0}\!\bra{0}^{\otimes R} \right) \right)$. Hence, we finish deriving the expression of $\hat{\rho}^{\mathsf{MR}}_n$.

\section{Deriving the \nisqrc{} quantum I/O map}

In this Supplementary Note, we will derive the I/O map of the \nisqrc{} framework, ultimately arriving at the results presented in Eq.\,(\ref{eq:Volterra}) of the main text.

\subsection{Technique of $u$-expansion and $\mathcal{R}_k$ and $\mathcal{P}_k$ superoperators}
\label{subsec:R_kandP_k}

In Supplementary Note \ref{app:meas_dynamics}, we have obtained concise formula Supplementary Equation \ref{apxeq:rhoMRn} for evaluating the infinitely-sampled readout features $x_j(n)$ under a general superoperator $\mathcal{U}(n)$ and a simple quantum measurement and reset scheme. However, the explicit dependence of these readout features on the input $u(n)$ - which defines the I/O map implemented by the \nisqrc{} scheme - is not yet apparent.

Uncovering this dependence requires addressing two complex, and in our framework, related issues. First, the I/O map is generally nonlinear in the input space. For example, in the Hamiltonian model we consider in main text, even if 
both the Hamiltonian encoding $\hat{H}(u)=\hat{H}_0 + u \cdot \hat{H}_1$ in Supplementary Equation \ref{eq:Hu=H0+uH1} and readouts $\langle \hat{M}_{\mathbf{w}} \rangle_{\hat{\rho}^{\mathsf{MR}}_n}$ are linear, the evolution defined by $\hat{U}(u) = e^{-i \tau \hat{H}(u)}$ will clearly lead to a nonlinear dependence on the inputs at every time step. Secondly, the map also extends over past inputs: the \nisqrc{} framework has memory. The dependence on past input history must be extricated by unraveling the recurrent structure of, for example, Supplementary Equation \ref{apxeq:rhoMRn}, necessitated by the multi-step nature of \nisqrc{} for temporal data processing. We will show that both these complications are addressable within a unified framework using a Volterra series description.

The key theoretical tool we employ to achieve this is referred to as the $u$-\textit{expansion}: an expansion of the superoperators governing dynamics in the \nisqrc{} framework, including measurement and reset, in powers of the input $u$. More precisely, we wish to expand the superoperators $\mathcal{U}(u)$ and $\mathcal{C}(u)$ in terms of the monomial $u^k$: 
\begin{align}
    \mathcal{U} (u) \hat{\rho}^{\mathsf{MR}} & = \sum_{k = 0}^{\infty} u^k \mathcal{R}_k \hat{\rho}^{\mathsf{MR}}, \label{eq:U=ukRk} \\
    \mathcal{C} (u) \hat{\rho}^{\mathsf{M }} & = \sum_{k = 0}^{\infty} u^k \mathcal{P}_k \hat{\rho}^{\mathsf{M }}. \label{eq:E=ukPk}
\end{align}
for some superoperators $\{\mathcal{R}_k\}$, $\{\mathcal{P}_k\}$ respectively. 

Regardless of the exact expression of $u$-expansion of the other dynamical superoperator, $\mathcal{C} (u) \hat{\rho}^{\mathsf{M}}$, the relationship between $\mathcal{U}(u)$ and $\mathcal{C}(u)$ means that the $u$-expansion of the latter may be directly derived from the $u$-expansion of the former. In particular,
\begin{align}
    \mathcal{C} (u) \hat{\rho}^{\mathsf{M}} & = \mathrm{Tr}_{\mathsf{R}} \! \left( \mathcal{U} (u) \! \left( \hat{\rho}^{\mathsf{M}} \otimes \ket{0} \! \bra{0}^{\otimes R} \right)  \right) = \sum_{k = 0}^{\infty} u^k \mathrm{Tr}_{\mathsf{R}} \! \left( \mathcal{R}_k \! \left( \hat{\rho}^{\mathsf{M}} \otimes \ket{0} \! \bra{0}^{\otimes R} \right)  \right).
\end{align}
where we have used Supplementary Equation \ref{eq:U=ukRk}. The final expression is exactly the desired form of Supplementary Equation \ref{eq:E=ukPk}, provided we make the identification
\begin{equation}
  \mathcal{P}_k \hat{\rho}^{\mathsf{M}} = \mathrm{Tr}_{\mathsf{R}} \! \left( \mathcal{R}_k \! \left( \hat{\rho}^{\mathsf{M}} \otimes \ket{0} \! \bra{0}^{\otimes R} \right)  \right) .
\end{equation}
If $k=0$, then in the main text we have already pointed out that the \textit{null-input superoperator} $\mathcal{P}_0 = \mathcal{C}(0)$ is a CPTP map. Furthermore, notice the expansion identity:
\begin{align}
    \mathcal{C}(u) \hat{\rho}^{\mathsf{M}} = \mathcal{P}_0 \hat{\rho}^{\mathsf{M}} + \sum_{k=1}^{\infty} u^k {\mathcal{P}_{k}} \hat{\rho}^{\mathsf{M}},
\end{align}
the trace-preserving nature, namely $\mathrm{Tr}(\mathcal{C}(u) \hat{\rho}^{\mathsf{M}}) = \mathrm{Tr}(\mathcal{P}_0 \hat{\rho}^{\mathsf{M}}) \equiv 1$, implies the tracelessness of $\mathcal{P}_k$ for all $k \geq 1$, i.e.
\begin{equation}
    \mathrm{Tr}(\mathcal{P}_k \hat{\rho}^{\mathsf{M}}) = 0. \label{eq:traceless}
\end{equation}
If we take $\mathcal{P}_k \hat{\rho}^{\mathsf{M}} = \sum_{\alpha = 1}^{4^M} c^{(k)}_{\alpha} \hat{\varrho}^{\mathsf{M}}_\alpha$, where $\hat{\varrho}^{\mathsf{M}}_\alpha$ are the eigenmatrices of superoperators $\mathcal{P}_0\hat{\varrho}^{\mathsf{M}}_\alpha = \lambda_\alpha \hat{\varrho}^{\mathsf{M}}_\alpha$. The decomposition coefficient $c^{(k)}_{1}$ is the most different one since its associated matrix $\hat{\varrho}^{\mathsf{M}}_1 = \hat{\rho}^{\mathsf{M}}_{\mathrm{FP}}$ will remain unchanged when applied by $\mathcal{P}_0$ while other modes decay to zero: $\lim_{n \to \infty} \mathcal{P}^{n}_0 \mathcal{P}_k \hat{\rho}_0^{\mathsf{M}} = c^{(k)}_{1} \hat{\rho}^{\mathsf{M}}_{\mathrm{FP}}$. As a result, 
\begin{align}
    c^{(k)}_{1} = c^{(k)}_{1} \mathrm{Tr}(\hat{\rho}^{\mathsf{M}}_{\mathrm{FP}}) = \lim_{n \to \infty} \mathrm{Tr}(\mathcal{P}^{n}_0 \mathcal{P}_k \hat{\rho}_0^{\mathsf{M}}) = \lim_{n \to \infty} \mathrm{Tr}(\mathcal{P}_k \hat{\rho}_0^{\mathsf{M}}) = 0. \label{eq:ck1=0}
\end{align}
Therefore, we conclude a very useful property that
\begin{align}
    \mathcal{P}_k \hat{\rho}^{\mathsf{M}} = \sum_{\alpha = 2}^{4^M} c^{(k)}_{\alpha} \hat{\varrho}^{\mathsf{M}}_\alpha \label{eq:alpha=2}
\end{align}
for any memory density matrix $\hat{\rho}^{\mathsf{M}}$ and any $k \geq 1$.

\subsection{$\mathcal{R}_k$ and $\mathcal{P}_k$ for linear Hamiltonian encoding scheme by regrouping the BCH formula}
\label{app:regroup_BCH}

We now evaluate the $u$-expansion of $\mathcal{U} (u) \hat{\rho}^{\mathsf{MR}} = e^{- i \tau \hat{H} (u)} \hat{\rho}^{\mathsf{MR}} e^{i \tau \hat{H} (u)}$. Central to this expansion is the Baker-Campbell-Hausdorff (BCH) formula, which allows us to write this expression in the series form
\begin{align}
    e^{- i \tau \hat{H} (u)} \hat{\rho}^{\mathsf{MR}} e^{i \tau \hat{H} (u)} = \sum_{q=0}^{\infty} \frac{(- i \tau)^q}{q!} [ \hat{H} (u), [ \cdots [ \hat{H} (u), \hat{\rho}^{\mathsf{MR}} ] \cdots ] ] 
    \label{apxeq:bchU}
\end{align}
Using the explicit form $\hat{H} (u) = \hat{H}_0 + u \hat{H}_1$, we can compute the superoperator coefficient of any term in the series:
\begin{align}
    \frac{(- i \tau)^1}{1!} : ~ [ \hat{H} (u), \hat{\rho}^{\mathsf{MR}} ] = & [ \hat{H}_0, \hat{\rho}^{\mathsf{MR}} ] + u^1 [ \hat{H}_1, \hat{\rho}^{\mathsf{MR}} ], \nonumber\\
    \frac{(- i \tau)^2}{2!} : ~ [ \hat{H} (u), [ \hat{H} (u), \hat{\rho}^{\mathsf{MR}} ] ] = & [ \hat{H}_0, [ \hat{H}_0, \hat{\rho}^{\mathsf{MR}} ] ] + u^1 \left( [ \hat{H}_0, [ \hat{H}_1, \hat{\rho}^{\mathsf{MR}} ] ] + [ \hat{H}_1, [ \hat{H}_0, \hat{\rho}^{\mathsf{MR}} ] ] \right) + u^2 [ \hat{H}_1, [ \hat{H}_1, \hat{\rho}^{\mathsf{MR}} ] ], \nonumber\\
    \frac{(- i \tau)^3}{3!} : ~ [ \hat{H} (u), [ \hat{H} (u), [ \hat{H} (u), \hat{\rho}^{\mathsf{MR}} ] ] ] = & [ \hat{H}_0, [ \hat{H}_0, [ \hat{H}_0, \hat{\rho}^{\mathsf{MR}} ] ] ] + \nonumber\\
    & + u^1 \left( [ \hat{H}_0, [ \hat{H}_0, [ \hat{H}_1, \hat{\rho}^{\mathsf{MR}} ] ] ] + [ \hat{H}_0, [ \hat{H}_1, [ \hat{H}_0, \hat{\rho}^{\mathsf{MR}} ] ] ] + [ \hat{H}_1, [ \hat{H}_0, [ \hat{H}_0, \hat{\rho}^{\mathsf{MR}} ] ] ] \right) \nonumber\\ 
    & + u^2 \left( [ \hat{H}_1, [ \hat{H}_1, [ \hat{H}_0, \hat{\rho}^{\mathsf{MR}} ] ] ] + [ \hat{H}_1, [ \hat{H}_0, [ \hat{H}_1, \hat{\rho}^{\mathsf{MR}} ] ] ] + [ \hat{H}_0, [ \hat{H}_1, [ \hat{H}_1, \hat{\rho}^{\mathsf{MR}} ] ] ] \right) \nonumber\\
    & + u^3 \left( [ \hat{H}_1, [ \hat{H}_1, [ \hat{H}_1, \hat{\rho}^{\mathsf{MR}} ] ] ] \right), \nonumber\\
    \vdots &  \nonumber
\end{align}
Note that each term in the series can be viewed as a series in $u^k$ instead. Furthermore, each appearance of $u \hat{H}_1$ in $\hat{H} (u)$ contributes exactly one factor of $u$. This allows us to determine the coefficient of $u^k$ in the $q$th term:
\begin{equation}
    u^k \times \frac{(- i \tau)^q}{q!} \sum_{\{ \hat{C}_1, \hat{C}_2, \cdots, \hat{C}_q \}} [ \hat{C}_1, [ \hat{C}_2, [ \cdots, [ \hat{C}_q, \hat{\rho}^{\mathsf{MR}} ] \cdots ] ] ]
\end{equation}
Here, the summation is over all $\left(\begin{array}{c}
  q\\
  k
\end{array}\right)$ possible combinations $\{ \hat{C}_1, \hat{C}_2, \cdots, \hat{C}_q \}$ which is an ordered set with $k$ instances of $\hat{H}_0$ and $(q - k)$ instances of $\hat{H}_1$. This expression allows us to regroup the BCH formula not by the parameter $q$ as in Supplementary Equation \ref{apxeq:bchU}, but by powers $u^k$ of the input. We therefore arrive at the desired form of Supplementary Equation \ref{eq:U=ukRk}, 
\begin{equation}
    \mathcal{U} (u) \hat{\rho}^{\mathsf{MR}} = \sum_{k = 0}^{\infty} u^k \mathcal{R}_k \hat{\rho}^{\mathsf{MR}},
\end{equation}
with
\begin{equation}
    \mathcal{R}_k \hat{\rho}^{\mathsf{MR}} = \sum_{q = k}^{\infty} \frac{(- i \tau)^q}{q!} \hspace{0mm} \sum_{\{ \hat{C}_1, \hat{C}_2, \cdots, \hat{C}_q \}} \hspace{0mm}  [ \hat{C}_1, [ \hat{C}_2, [ \cdots, [ \hat{C}_q, \hat{\rho}^{\mathsf{MR}} ] \cdots ] ] ] . \label{eq:R_k}
\end{equation}

\subsection{Functional I/O map: time-invariance and Volterra kernels}
\label{app:volterra_kernel_derive}

Our work in the previous subsection allows us to express the action of individual superoperators $\mathcal{U}(u)$ and $\mathcal{C}(u)$ on a general $\hat{\rho}^{\mathsf{MR}}$ as a $u$-expansion at every time step. The dynamical map defined by our time-dependent \nisqrc{} framework involves the repeated application of these superoperators for distinct inputs $u_{n}$, so that the output at time step $n$ may have a complicated dependence on prior inputs $u_{\leq n}$. We are now in a position to extract this dependence explicitly. To do so, we simply substitute our $u$-expansions for the superoperators $\mathcal{U} (u)$ and $\mathcal{C} (u)$ into the evolution equation Supplementary Equation \ref{apxeq:rhoMRn} defining $\hat{\rho}^{\mathsf{MR}}_n$ at an arbitrary time step $n$, i.e. $\hat{\rho}^{\mathsf{MR}}_n =\mathcal{U} (u_n) \left( \left( \mathcal{C} (u_{n - 1}) \cdots \mathcal{C} (u_1) \hat{\rho}^{\mathsf{M}}_0 \right) \otimes \ket{0}\!\bra{0}^{\otimes R} \right)$. Then, the density matrices at time step $n$ attain the formal expression:
\begin{align}
    \hat{\rho}^{\mathsf{MR}}_n = \hspace{-2mm} \sum_{k_1, \cdots, k_n = 0}^{\infty} \hspace{-2mm} u_1^{k_1} \cdots u_{n-1}^{k_{n-1}} u_n^{k_n} \times \mathcal{R}_{k_n} \!\left( \left( \mathcal{P}_{k_{n-1}} \cdots \mathcal{P}_{k_1} \hat{\rho}^{\mathsf{M}}_0 \right) \otimes \ket{0}\!\bra{0}^{\otimes R} \right) \label{eq:vanilla_rhon}
\end{align}

Before evaluating the readout features $x_j(n)$, we need to simplify Supplementary Equation \ref{eq:vanilla_rhon} as much as possible. The starting point is first looking at the simplest contribution from term $u_{n - 1}$ to $\hat{\rho}^{\mathsf{MR}}_n$ (namely the one-step backwards linear contribution). This means that we can let $k_1 = \cdots = k_{n - 2} = k_n = 0$ and $k_{n - 1} = 1$. The associated prefactor is
\begin{equation}
    \mathcal{R}_0 \left( \left( \mathcal{P}_1 \mathcal{P}^{n - 2}_0 \hat{\rho}^{\mathsf{M}}_0 \right) \otimes \ket{0}\!\bra{0}^{\otimes R} \right), \label{eq:prefactor1}
\end{equation}
Similarly, analyzing contribution from term $u_n$ to $\hat{\rho}^{\mathsf{MR}}_{n + 1}$ (that is, let $k_1 = \cdots = k_{n - 1} = k_{n + 1} = 0$ and $k_n = 1$) gives associated prefactor
\begin{equation}
    \mathcal{R}_0 \left( \left( \mathcal{P}_1 \mathcal{P}^{n - 1}_0 \hat{\rho}^{\mathsf{M}}_0 \right) \otimes \ket{0}\!\bra{0}^{\otimes R} \right). \label{eq:prefactor2}
\end{equation}
In principle, $\mathcal{P}^{n - 2}_0 \hat{\rho}^{\mathsf{M}}_0 \neq \mathcal{P}^{n - 1}_0 \hat{\rho}^{\mathsf{M}}_0$ and therefore term Supplementary Equation \ref{eq:prefactor1} and Supplementary Equation \ref{eq:prefactor2} are analytically different. However, with the existence of fixed point state
\begin{equation}
    \lim_{n \rightarrow \infty} \mathcal{P}^n_0 \hat{\rho}^{\mathsf{M}}_0 = \hat{\rho}^{\mathsf{M}}_{\mathrm{FP}},
\end{equation}
it ensures the approximation
\begin{equation}
    \mathcal{P}^{n - 2}_0 \hat{\rho}^{\mathsf{M}}_0 \approx \hat{\rho}^{\mathsf{M}}_{\mathrm{FP}} \approx \mathcal{P}^{n - 1}_0 \hat{\rho}^{\mathsf{M}}_0,
\end{equation}
and hence Supplementary Equation \ref{eq:prefactor1} and Supplementary Equation \ref{eq:prefactor2} are asymptotically the same. Such property is usually referred as (asymptotic) \textit{time-invariance}. In fact, we can further weaken this requirement that all peripheral spectrum $\lambda_\alpha$ (namely those eigenvalue with magnitude $|\lambda_\alpha|=1$) are $\lambda_\alpha=1$. For example, for a fully connected quantum reservoir with $M+R$ qubits, if $J_{i,i'}$ are constant for every coupling pair and $\eta^x_i, \eta^z_i$ are also constant for every qubit, then the numerical results show that the fixed points of $\hat{\rho}^{\mathsf{M}}_0$ will have a degeneracy of \textit{Catalan numbers} $\frac{(2M)!}{M!(M+1)!}$. In this case, the fixed point $\lim_{n \to \infty} \mathcal{P}^{n}_0 \hat{\rho}^{\mathsf{M}}_0 = \hat{\rho}^{\mathsf{M}}_{\mathrm{FP}}$ still exists but will depend on initial state $\hat{\rho}^{\mathsf{M}}_0$. 

The above calculation works for any contribution terms in $\hat{\rho}^{\mathsf{MR}}_n$. This establishes all analytical expressions of Volterra series kernels. The leading order kernels can be written down compactly:
\begin{itemize}
    \item The zero-th order Volterra kernel:
    \begin{align}
        h_0^{(j)} = \mathrm{Tr} \! \left( \hat{M}_j \mathcal{R}_0 \! \left( \hat{\rho}^{\mathsf{M}}_{\mathrm{FP}} \otimes \ket{0} \! \bra{0}^{\otimes R} \right) \right),
        \label{apxeq:kernel0}
    \end{align}
    \item The first order Volterra kernel ($n_1 \geq 0$):
    \begin{align}
        h_1^{(j)} (n_1) = \left\{\begin{array}{ll}
            \mathrm{Tr} \! \left( \hat{M}_j \mathcal{R}_1 \! \left( \hat{\rho}^{\mathsf{M}}_{\mathrm{FP}} \otimes \ket{0} \! \bra{0}^{\otimes R} \right) \right), & \text{if } n_1 = 0,\\
            \mathrm{Tr} \! \left( \hat{M}_j \mathcal{R}_0 \! \left( \left( \mathcal{P}^{n_1 - 1}_0 \mathcal{P}_1 \hat{\rho}^{\mathsf{M}}_{\mathrm{FP}} \right) \otimes \ket{0} \! \bra{0}^{\otimes R} \right) \right), & \text{if } n_1 \neq 0,
      \end{array}\right. 
      \label{apxeq:kernel1}
    \end{align}
    \item And the second order Volterra kernel ($n_2 \geq n_1 \geq 0$):
    \begin{align}
        h_2^{(j)} (n_1, n_2) =
        \left\{\begin{array}{ll}
            \mathrm{Tr} \! \left( \hat{M}_j \mathcal{R}_2 \! \left( \hat{\rho}^{\mathsf{M}}_{\mathrm{FP}} \otimes \ket{0} \! \bra{0}^{\otimes R} \right) \right), & \text{if } n_1 = 0, n_2 = 0,\\
            \mathrm{Tr} \! \left( \hat{M}_j \mathcal{R}_1 \! \left( \left( \mathcal{P}^{n_2 - 1}_0 \mathcal{P}_1 \hat{\rho}^{\mathsf{M}}_{\mathrm{FP}} \right) \otimes \ket{0} \! \bra{0}^{\otimes R} \right) \right), & \text{if } n_1 = 0, n_2 > 0,\\
            \mathrm{Tr} \! \left( \hat{M}_j \mathcal{R}_0 \! \left( \left( \mathcal{P}^{n_2 - 1}_0 \mathcal{P}_2 \hat{\rho}^{\mathsf{M}}_{\mathrm{FP}} \right) \otimes \ket{0} \! \bra{0}^{\otimes R} \right) \right), & \text{if } n_1 = n_2 > 0,\\
            \mathrm{Tr} \! \left( \hat{M}_j \mathcal{R}_0  \!\left( \left( \mathcal{P}^{n_1 - 1}_0 \mathcal{P}_1 \mathcal{P}^{n_2 - n_1 - 1}_0 \mathcal{P}_1 \hat{\rho}^{\mathsf{M}}_{\mathrm{FP}} \right) \otimes \ket{0} \! \bra{0}^{\otimes R} \right) \right), & \text{if } 0 < n_1 < n_2 .
        \end{array}\right. 
        \label{apxeq:kernel2}
    \end{align}
\end{itemize}
These kernel expressions show that if the reservoir output nontrivially depends on the history, then $\mathcal{P}_k  \hat{\rho}^{\mathsf{M}}_{\mathrm{FP}} \neq 0$ for some $k \geq 1$. Equivalently, if $\mathcal{P}_k  \hat{\rho}^{\mathsf{M}}_{\mathrm{FP}} = 0$ for all $k \geq 1$, then $h^{(j)}_k(n_1, n_2, \cdots, n_k) \neq 0$ only if $n_1=n_2=\cdots=n_k=0$. 

We emphasize that even though $h^{(j)}_k(n_1, n_2, \cdots, n_k)$ (e.g., Supplementary Equation \ref{apxeq:kernel0}-\ref{apxeq:kernel2} are kernels of real values $x_j(n)$, these kernels all take the form of $\mathrm{Tr}(\hat{M}_j \cdot)$, where ``$\cdot$'' are always quantum operators which expand $\hat{\rho}^{\mathsf{MR}}_n$. Therefore, it is intuitive to write these quantum operators into
\begin{align}
    \hat{\rho}^{\mathsf{MR}}_n = \sum_{k = 0}^{\infty} \sum_{n_1 = 0}^{\infty} \cdots \hspace{-3mm} \sum_{n_k = n_{k-1}}^{\infty} \hspace{-3mm} \hat{h}_k (n_1, \cdots, n_k) \prod_{\kappa = 1}^{k} u_{n - n_{\kappa}}.
\end{align} 
Those quantum operators $\hat{h}_k$ are the central objects in the $u$-expansion and all classical kernels in Eq.\,(\ref{eq:Volterra}) are $h^{(j)}_k = \mathrm{Tr}(\hat{M}_j \hat{h}_k)$, justifying the nomenclature of Quantum Volterra Theory used for the entire framework in the main text.

We note that as we proved in Supplementary Equation \ref{eq:IL2L}, $\lim_{n \rightarrow \infty} \hat{\rho}_n^{\mathsf{MR}} = \frac{\hat{I}^{\otimes L}}{2^L}$. This can also be understood through the Volterra expansion. Recall Supplementary Equation \ref{eq:UMrho}, i.e. $\hat{\rho}_n^{\mathsf{MR}} = \mathcal{U} (u_n) \mathcal{M} \cdots \mathcal{U} (u_2) \mathcal{M} \mathcal{U} (u_1) \rho_0^{\mathsf{MR}}$. By plugging Supplementary Equation \ref{eq:U=ukRk} and Supplementary Equation \ref{eq:E=ukPk}, we get 
\begin{align}
    \hat{\rho}^{\mathsf{MR}}_n = \hspace{-3mm} \sum_{k_1, k_2, \cdots, k_n = 0}^{\infty} \hspace{-2mm} u_1^{k_1} u_2^{k_2} \cdots u_n^{k_n} \mathcal{R}_{k_n} \mathcal{M} \mathcal{R}_{k_{n-1}} \cdots \mathcal{M} \mathcal{R}_{k_2} \mathcal{M} \mathcal{R}_{k_1} \hat{\rho}^{\mathsf{MR}}_0
\end{align}
All $\mathcal{P}_k \hat{\rho}^{\mathsf{M}}$ in previous Volterra analysis must be replaced with $\mathcal{M} \mathcal{R}_k \hat{\rho}^{\mathsf{MR}}$. However, $\mathcal{M} \mathcal{R}_0 \hat{\rho}^{\mathsf{MR}}_{\mathrm{FP}} = \hat{\rho}^{\mathsf{MR}}_{\mathrm{FP}}$ implies $\hat{\rho}^{\mathsf{MR}}_{\mathrm{FP}} = \frac{\hat{I}^{\otimes L}}{2^L}$, and thus all Volterra kernels must vanish, since the identity makes all commutator terms in Supplementary Equation \ref{eq:R_k} vanish exactly. This reproduces the null response of a \nisqrc{} architecture in the absence of the reset operation.

\subsection{$u$-expansion and Volterra kernels for dissipative quantum systems}
\label{app:u_expension_dissipation}

Thus far, we have demonstrated how the $u$-expansion can be performed for a CPTP map without explicit dissipative evolution. In this subsection, we extend this analysis to account for dissipative quantum systems, as is relevant for practical \nisqrc{} implementations.

In particular, we wish to now consider the evolution governed by the general CPTP map $e^{\tau \mathcal{L} (u)} \hat{\rho}^{\mathsf{MR}}$, where $\mathcal{L}$ is the Liouvillian superoperator, for example of the type introduced in Eq.\,(\ref{eq:dissipation}) of the main text. We first note that the BCH formula of Supplementary Equation \ref{apxeq:bchU} can be rewritten compactly in the dissipation free case by first introducing the \textit{adjoint action} $[\hat{X}, \hat{Y}] = \mathrm{ad}_{\hat{X}} \hat{Y}$ for arbitrary matrices $\hat{X}, \hat{Y}$. With this notation, the BCH formula becomes:
\begin{equation}
    e^{- i \tau \hat{H}(u)} \hat{\rho} e^{i \tau \hat{H}(u)} = e^{- i \tau [\hat{H}(u), \cdot]} \hat{\rho} = \sum_{q = 0}^{\infty} \frac{(- i \tau)^q}{q!} \mathrm{ad}^q_{\hat{H}(u)} \hat{\rho} .
\end{equation}
In presence of dissipation, the adjoint action allows us to write the operation of the Liouvillian $\mathcal{L}(u)$, $e^{\tau \mathcal{L} (u)} \hat{\rho}^{\mathsf{MR}}$, in the form:
\begin{align}
    e^{\tau \mathcal{L} (u)} \hat{\rho}^{\mathsf{MR}} = & e^{- i \tau \left( \left( \mathrm{ad}_{\hat{H}_0} + i\mathcal{D}_{\mathrm{T}} \right) + u~\mathrm{ad}_{\hat{H}_1} \right)} \hat{\rho}^{\mathsf{MR}} \nonumber\\
    = & \hat{\rho}^{\mathsf{MR}} + \frac{(- i \tau)^1}{1!} \left( \left( \mathrm{ad}_{\hat{H}_0} + i\mathcal{D}_{\mathrm{T}} \right) + u~\mathrm{ad}_{\hat{H}_1} \right) \hat{\rho}^{\mathsf{MR}} + \frac{(- i \tau)^2}{2!} \left( \left( \mathrm{ad}_{\hat{H}_0} + i\mathcal{D}_{\mathrm{T}} \right) + u~\mathrm{ad}_{\hat{H}_1} \right)^2 \hat{\rho}^{\mathsf{MR}} + \cdots . 
\end{align}
where we have also used the explicit form of $\hat{H}(u) = \hat{H}_0 + u \hat{H}_1$, and where $\mathcal{D}_{\rm T}$ describes $T_1$ decay of all qubits in the QRC with a rate $\gamma$, see Eq.\,(\ref{eq:dissipation}) of the main text.

Based on this formalism, we are now able to read off the $u$-expansion for the CPTP map $e^{\tau \mathcal{L} (u)} \hat{\rho}^{\mathsf{MR}}$ by regrouping modified BCH formula:
\begin{equation}
    e^{\tau \mathcal{L} (u)} \hat{\rho}^{\mathsf{MR}} = \sum_{k = 0}^{\infty} u^k \mathcal{Q}_k \hat{\rho}^{\mathsf{MR}},
\end{equation}
where the superoperators are defined as
\begin{align}
    \mathcal{Q}_0 \hat{\rho}^{\mathsf{MR}} & = \hat{\rho}^{\mathsf{MR}} - i \tau \left( \mathrm{ad}_{\hat{H}_0} + i\mathcal{D}_{\mathrm{T}} \right) \hat{\rho}^{\mathsf{MR}} - \frac{\tau^2}{2!} \left( \mathrm{ad}_{\hat{H}_0} + i\mathcal{D}_{\mathrm{T}} \right)^2 \hat{\rho}^{\mathsf{MR}} + \cdots \nonumber\\
    & = \hat{\rho}^{\mathsf{MR}} - i \tau \left( [ \hat{H}_0, \hat{\rho}^{\mathsf{MR}} ] +  {i\mathcal{D}_{\mathrm{T}} \hat{\rho}^{\mathsf{MR}}} \right) - \frac{\tau^2}{2!} \left( 
        [ \hat{H}_0, [ \hat{H}_0, \hat{\rho}^{\mathsf{MR}} ] ] 
        + i \! [ \hat{H}_0, \mathcal{D}_{\mathrm{T}} \hat{\rho}^{\mathsf{MR}} ] 
        + i \mathcal{D}_{\mathrm{T}} [ \hat{H}_0, \hat{\rho}^{\mathsf{MR}} ] 
        - \mathcal{D}^2_{\mathrm{T}} \hat{\rho}^{\mathsf{MR}} 
    \right) + \cdots \nonumber\\
    & \equiv e^{\tau \mathcal{L} (0)} \hat{\rho}^{\mathsf{MR}}, \nonumber\\ 
    \mathcal{Q}_1 \hat{\rho}^{\mathsf{MR}} & = - i \tau \mathrm{ad}_{\hat{H}_1} \hat{\rho}^{\mathsf{MR}} - \frac{\tau^2}{2!} \left( \left( \mathrm{ad}_{\hat{H}_0} + i\mathcal{D}_{\mathrm{T}} \right) \mathrm{ad}_{\hat{H}_1} + \mathrm{ad}_{\hat{H}_1}  \! \left( \mathrm{ad}_{\hat{H}_0} + i\mathcal{D}_{\mathrm{T}} \right) \right) \hat{\rho}^{\mathsf{MR}} + \cdots, \nonumber\\
    & = - i \tau [ \hat{H}_1, \hat{\rho}^{\mathsf{MR}} ] - \frac{\tau^2}{2!} \left( [ \hat{H}_0, [ \hat{H}_1, \hat{\rho}^{\mathsf{MR}} ] ] + [ \hat{H}_1, [ \hat{H}_0, \hat{\rho}^{\mathsf{MR}} ] ] + {i\mathcal{D}_{\mathrm{T}}  \! [ \hat{H}_1, \hat{\rho}^{\mathsf{MR}} ] + i [ \hat{H}_1, \mathcal{D}_{\mathrm{T}} \hat{\rho}^{\mathsf{MR}} ]} \right) + \cdots, \nonumber\\
    \mathcal{Q}_2 \hat{\rho}^{\mathsf{MR}} & = - \frac{\tau^2}{2!} \mathrm{ad}_{\hat{H}_1}^2 \hat{\rho}^{\mathsf{MR}} + \cdots = - \frac{\tau^2}{2!} [ \hat{H}_1, [ \hat{H}_1, \hat{\rho}^{\mathsf{MR}} ] ] + \cdots. \nonumber\\
    & \hspace{2mm} \vdots  \nonumber
\end{align}

The knowledge of superoperators $\{\mathcal{Q}_k\}$ therefore allows us to compute the Volterra kernels for \nisqrc{} in the presence of dissipation. Some numerical simulations of the first- and second-order kernels are shown in Supplementary Figure \ref{fig:Kernel-2}, with increasing decay rate $\gamma$ (for other QRC parameters, see caption). We see that dissipation can reduce the amplitude of the QRC response to the input - governed by the amplitude of the kernels - in particular to past inputs indicated by increasing values of $n_1, n_2$. Hence dissipation can reduce the memory of the \nisqrc{} framework. However, even for modest amounts of dissipation smaller than the strength of Hamiltonian terms, the kernels are certainly far from trivial, retaining their qualitative features with a non-zero memory term. This indicates the applicability of the \nisqrc{} framework to contemporary dissipative quantum systems used as QRCs.


\begin{figure*}[t]
    \centering
    \includegraphics[width=\textwidth]{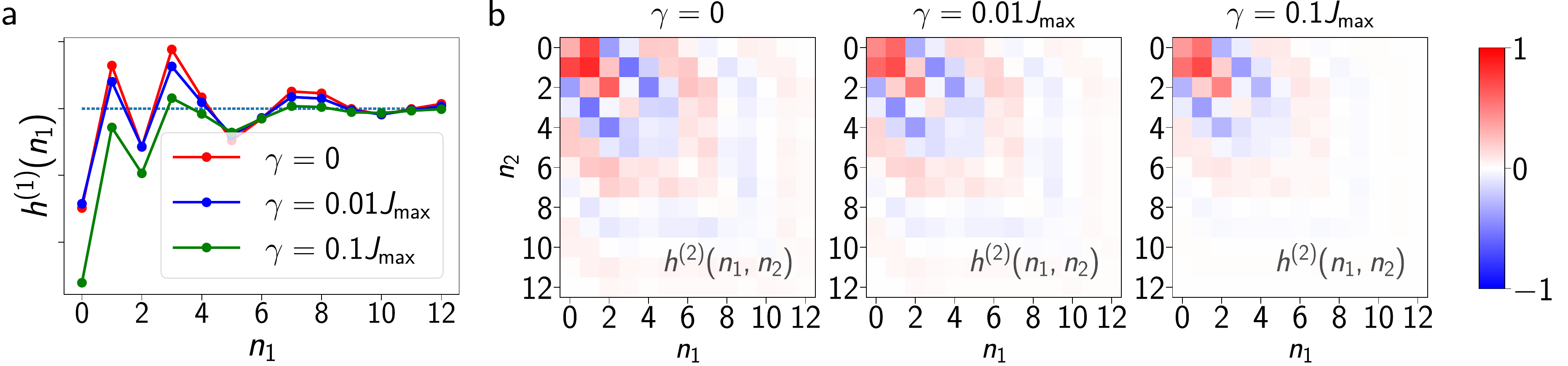}
    \caption{The first and second order Volterra Kernel example in a $(2+1)$-qubit quantum reservoir with fundamental decay $\gamma$. The parameters are chosen to be constant $J_{i,i'} = \eta^x_i = \eta^z_i = 1.3$ (in unit $1/\tau$) for simplicity. (The same as Fig.\,\ref{fig:nM_T1} of the main text). (a) The first order kernel $h^{(1)}_1(n_1)$, with decay rate $\gamma = 0$ (no decay, in red), $ 0.01J_{\mathrm{max}}$ (in blue), $0.1J_{\mathrm{max}}$ (in green). (b) The second order kernel $h^{(1)}_2(n_1, n_1)$, with decay rate $\gamma = 0$ (no decay, left), $ 0.01J_{\mathrm{max}}$ (middle), $0.1J_{\mathrm{max}}$ (right). The first and second order kernel without decay is exactly the kernel in Fig.\,\ref{fig:nM_T1}(a) of the main text.}
    \label{fig:Kernel-2}
\end{figure*}


\section{Fading memory modes}
\label{app:memory_modes}
For any $k \geq 1$, we define for each $\alpha' \in \mathbb{N}$
\begin{equation}
    \mathcal{P}_k \hat{\varrho}^{\mathsf{M}}_{\alpha'} = \sum_{\alpha = 2}^{4^M} c_{\alpha \alpha'}^{(k)} \hat{\varrho}^{\mathsf{M}}_{\alpha} .
\end{equation}
Notice that $c_{1 \alpha'}^{(k)} \equiv 0$ for any $k \geq 1$ due to the tracelessness of $\mathcal{P}_k$ (recall $\hat{\varrho}^{\mathsf{M}}_{1} = \hat{\rho}^{\mathsf{M}}_{\mathrm{FP}}$ by definition), thus the summation begins with $\alpha = 2$. Contributions from $(u_{n - n_1}, u_{n - n_2}, \cdots, u_{n - n_P})$, where $0 < n_1 < n_2 < \cdots < n_P$, is given by
\begin{align}
    & \sum_{k_1, \cdots, k_P = 1}^{\infty} h^{(j)}_{k_1+\cdots+k_P}(n^{\otimes k_1}_{1}, \cdots, n^{\otimes k_P}_{P}) \times u_{n - n_1}^{k_1} \cdots u_{n - n_P}^{k_P} \nonumber\\
    = & \sum_{k_1, \cdots, k_P = 1}^{\infty} \mathrm{Tr} \! \left( \hat{M}_j \mathcal{R}_0 \! \left( \mathcal{P}^{n_1 - 1}_0 \mathcal{P}_{k_1} \cdots \mathcal{P}^{n_P - n_{P - 1} - 1}_0 \mathcal{P}_{k_P} \hat{\rho}^{\mathsf{M}}_{\mathrm{FP}} \otimes \ket{0} \! \bra{0}^{\otimes R} \right) \right) \times u_{n - n_1}^{k_1} \cdots u_{n - n_P}^{k_P} \nonumber\\
    = & \sum_{k_1, \cdots, k_P = 1}^{\infty} \mathrm{Tr} \! \left( \hat{M}_j \! \mathcal{R}_0 \left( \mathcal{P}^{n_1 - 1}_0 \mathcal{P}_{k_1} \cdots \mathcal{P}^{n_P - n_{P - 1} - 1}_0 \left( \sum_{\alpha_P = 2}^{4^M} c_{\alpha_P 1}^{(k_P)} \hat{\varrho}^{\mathsf{M}}_{\alpha_P} \right) \otimes \ket{0} \! \bra{0}^{\otimes R} \right) \right) \times u_{n - n_1}^{k_1} \cdots u_{n - n_P}^{k_P} \nonumber\\
    \vdots & \nonumber\\
    = & \sum_{k_1, \cdots, k_P = 1}^{\infty} \sum_{\alpha_1, \cdots, \alpha_P = 2}^{4^M} \mathrm{Tr} \! \left( \hat{M}_j \mathcal{R}_0 \! \left( \lambda^{n_1 - 1}_{\alpha_1} c_{\alpha_1 \alpha_2}^{(k_1)} \cdots \lambda^{n_P - n_{P - 1} - 1}_{\alpha_P} c_{\alpha_P 1}^{(k_P)} \hat{\varrho}^{\mathsf{M}}_{\alpha_1} \otimes \ket{0} \! \bra{0}^{\otimes R} \right) \right) \times u_{n - n_1}^{k_1} \cdots u_{n - n_P}^{k_P} \nonumber\\
    = & \sum_{\alpha_1, \cdots, \alpha_P = 2}^{4^M} \lambda^{n_1 - 1}_{\alpha_1} \cdots \lambda^{n_P - 1}_{\alpha_P} \sum_{k_1, \cdots, k_P = 1}^{\infty} c_{\alpha_1 \alpha_2}^{(k_1)} \cdots c_{\alpha_P 1}^{(k_P)} \mathrm{Tr} \! \left( \hat{M}_j \mathcal{R}_0 \! \left( \hat{\varrho}^{\mathsf{M}}_{\alpha_1} \otimes \ket{0} \! \bra{0}^{\otimes v} \right) \right) \times u_{n - n_1}^{k_1} \cdots u_{n - n_P}^{k_P} . 
\end{align}
Namely, we can decompose the contributions from $(u_{n - n_1}, u_{n - n_2}, \cdots, u_{n - n_P})$ to $x_j (n)$ into $(4^M - 1)^P$ memory modes of internal features:
\begin{align}
    x_j (n) = & \sum_{\alpha_1, \alpha_2, \cdots, \alpha_P = 2}^{4^M} \nu^{(j)}_{\alpha_1} \lambda^{n_1 - 1}_{\alpha_1} \lambda^{n_2 - n_1 - 1}_{\alpha_2} \cdots \lambda^{n_P - n_{P - 1} - 1}_{\alpha_P} \nonumber\\
    & \hspace{20mm} \times F_{\alpha_1, \alpha_2, \cdots, \alpha_P} (u_{n - n_1}, u_{n - n_2}, \cdots, u_{n - n_P}) + \cdots \label{eq:xj_cross_contribution}
\end{align}
where the cross-step internal features
\begin{align}
    F_{\alpha_1, \alpha_2, \cdots, \alpha_P} (u_{n - n_1}, u_{n - n_2}, \cdots, u_{n - n_P}) = \sum_{k_1, k_2 \cdots, k_P = 1}^{\infty} c_{\alpha_1 \alpha_2}^{(k_1)} c_{\alpha_2 \alpha_3}^{(k_2)} \cdots c_{\alpha_P 1}^{(k_P)} u_{n - n_1}^{k_1} u_{n - n_2}^{k_2} \cdots u_{n - n_P}^{k_P}.
\end{align}


\begin{figure}[t]
    \centering
    \includegraphics[width = 0.8\columnwidth]{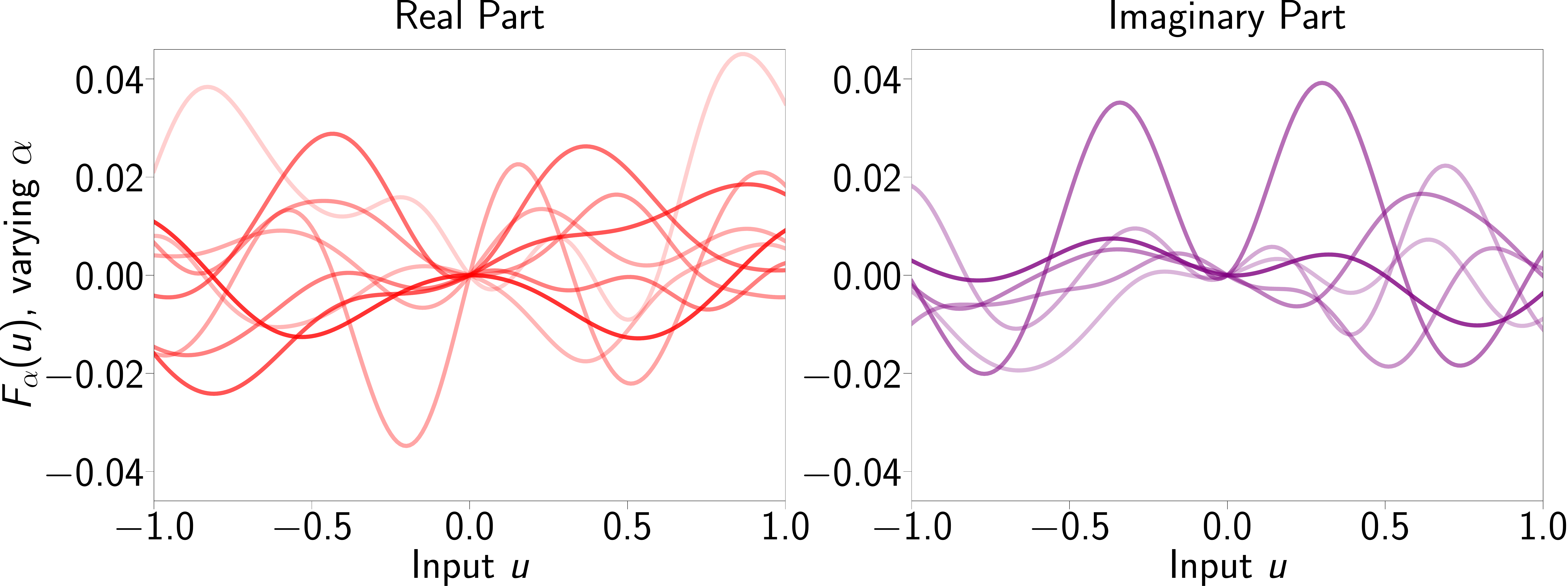}
    \caption{$4^M-1 = 15$ internal features $F_\alpha(u)$ in a $(2+1)$-qubit QRC. The hyperparameters are $(J_{\mathrm{max}}; \eta^x, \varepsilon^x_{\mathrm{rms}}; \eta^z, \varepsilon^z_{\mathrm{rms}}) = (1; 3, 1; 4, 2)$ in unit $1/\tau$. $F_\alpha(u)$ is potentially a complex-valued function. The eigenvalues of $\mathcal{P}_0$ appears in pair: $\lambda_\alpha$ being an eigenvalue implies $\lambda^*_\alpha$ also being an eigenvalue. Therefore, both function $F_\alpha(u)$ and its conjugate $F_\alpha(u)^*$ are internal features. So the of $4^M-1$ internal feature functions contains exactly $4^M-1$ independent real-function. They are plotted separately in sense of of real part (red solid lines) and imaginary part (purple solid lines). The darker the line is, the larger the corresponding eigenvalue norm $|\lambda_\alpha|$ is and the slower this internal feature fades. }
    \label{fig:Internal_Features}
\end{figure}


Thanks to the \textit{fading memory} property, namely that $\lambda^n_{\alpha}$ converges to zero if $\alpha \geq 2$, the more history steps one monomial term in Volrerra series Eq.\,(\ref{eq:Volterra}) involves, the less it contributes to the current-time readout features $x_j(n)$. Therefore, it will be illustrative for this Supplementary Note to mostly be concerned with a single past time step's contribution. To be more specific, if we focus on the contribution from $u_{n-p}$ to $x_j(n)$ (where $p \geq 1$). For this history record contribution, 
\begin{align}
    \sum_{k=1}^{\infty} \! h_k^{(j)}\!(p^{\otimes k}) \, u^k_{n - p} = \sum_{\alpha = 2}^{4^M} \nu^{(j)}_{\alpha} \lambda^{p - 1}_\alpha F_\alpha(u_{n-p}).
\end{align}
Each coefficient $\nu^{(j)}_{\alpha} = \mathrm{Tr} \! \left( \hat{M}_j \mathcal{R}_0 \! \left( \hat{\varrho}^{\mathsf{M}}_\alpha \otimes \ket{0} \! \bra{0}^{\otimes R} \right) \right)$ characterizes a different observable $\hat{M}_j$'s response to different \textit{internal features} $F_\alpha (u)$ where
\begin{align}
    F_\alpha (u) = \sum_{k = 1}^{\infty} c^{(k)}_{\alpha 1} u^k . 
    \label{appeq:falpha}
\end{align}
Especially, if $\alpha=1$, then $c^{(k)}_{\alpha 1} = 0$ for any $k \geq 1$, according to Supplementary Equation \ref{eq:ck1=0}. That is why the summation over $\alpha$ starts from $\alpha=2$, and it only gives us $4^M-1$ internal features (see Supplementary Figure \ref{fig:Internal_Features} as an example).

\section{Relation between functional-independence and Jacobian rank}
\label{app:Function-independence}

In this Supplementary Note we analyze the functional-independence of readout features in the \nisqrc{} framework. Assuming a finite-dimensional input space $\UI = (u_1, u_2, \cdots, u_N)$, then \nisqrc{} readout features define $K$ finite-dimensional functions (assuming $K \leq N$), $x_k (u_1, u_2, \cdots, u_N)$, $k\in\{0,\cdots,K-1\}$. An important question is whether these $K$ functions are in fact functionally-independent from one another, since their inter-dependence can impose a limitation on their usefulness for functional approximation using the \nisqrc{} framework.

If the $K$ functions are functionally-dependent, namely there exists some $K$-variate function $G$ such that:
\begin{equation}
    G (x_0 (\UI), x_1 (\UI), \cdots, x_{K-1} (\UI) ) \equiv 0.
\end{equation}
Take gradients
\begin{equation}
    \left(\begin{array}{c}
        \frac{\partial G}{\partial u_1} (\UI) \\
        \frac{\partial G}{\partial u_2} (\UI) \\
        \vdots \\
        \frac{\partial G}{\partial u_N} (\UI) 
    \end{array}\right) = 
    \left(\begin{array}{cccc}
        \frac{\partial x_0}{\partial u_1} (\UI) & \frac{\partial x_1}{\partial u_1} (\UI) & \cdots & \frac{\partial x_{K-1}}{\partial u_1} (\UI) \\
        \frac{\partial x_0}{\partial u_2} (\UI) & \frac{\partial x_1}{\partial u_2} (\UI) & \cdots & \frac{\partial x_{K-1}}{\partial u_2} (\UI) \\
        \vdots & \vdots & \ddots & \vdots \\
        \frac{\partial x_0}{\partial u_N} (\UI) & \frac{\partial x_1}{\partial u_N} (\UI) & \cdots & \frac{\partial x_{K-1}}{\partial u_N} (\UI)
    \end{array}\right) 
    \left(\begin{array}{c}
        \frac{\partial G}{\partial x_0} (x_0, x_1, \cdots, x_{K-1}) \\
        \frac{\partial G}{\partial x_1} (x_0, x_1, \cdots, x_{K-1}) \\
        \vdots \\
        \frac{\partial G}{\partial x_{K-1}} (x_0, x_1, \cdots, x_{K-1}) 
    \end{array}\right)
    = 0,
\end{equation}
then gradients $\nabla_{\UI} x_0 (\UI)$, $\nabla_{\UI} x_1 (\UI)$, $\cdots, \nabla_{\UI} x_{K-1} (\UI)$ must be linearly dependent at all points. Therefore, if $\{ x_j (\UI) \}_{j \in [K]}$ are functionally-dependent, then the gradients $\{ \nabla_{\UI} x_j (\UI) \}_{j \in [K]}$ must be linearly-dependent. Equivalently, it suffices to prove the functional-independence of $\{ x_j (\UI) \}_{j \in [K]}$ by showing that $\{ \nabla_{\UI} x_j (\UI) \}_{j \in [K]}$ are linearly-independent at almost all points $\UI$.

Now we argue by contradiction that $K-1$ gradients of readout features $x_j (n) = \mathrm{Tr} \left( \hat{M}_j \hat{\rho}_n^{\mathsf{MR}} \right)$ are functionally-independent if there is no particular symmetry in the reservoir. We first select $\{\hat{M}_j\}$ as the moment representation to remove the trivial functional dependence that their summation is constant. 
Suppose there exists coefficients $c_1, c_2, \cdots, c_{K-1}$ such that $\sum^{K-1}_{j = 1} c_j \nabla \mathcal{F}_j (u_{\leq n}) = 0$. Notice that
\begin{equation}
    \frac{\partial x_j}{\partial u_{n - p}} = \mathrm{Tr} \left( \hat{M}_j
    \frac{\partial \hat{\rho}_n^{\mathsf{MR}}}{\partial u_{n - p}} \right),
\end{equation}
then $\sum^{K-1}_{j = 1} c_j \nabla \mathcal{F}_j (u_{\leq n}) = 0$ implies that
\begin{equation}
    \sum^{K-1}_{j = 1} c_j \mathrm{Tr} \left( \hat{M}_j \frac{\partial \hat{\rho}_n^{\mathsf{MR}}}{\partial u_{n - p}} \right) = \mathrm{Tr} \left( \left( \sum^{K-1}_{j = 1} c_j \hat{M}_j \right) \frac{\partial \hat{\rho}_n^{\mathsf{MR}}}{\partial u_{n - p}} \right) \equiv 0,
\end{equation}
for all non-negative integer $p \in \mathbb{N}$. For generic input sequence $\{ u_{- \infty}, \cdots, u_{n - 1}, u_n \}$, there doesn't exists such observable $\sum^{K-1}_{j = 1} c_j \hat{M}_j$ such that expectations of $\frac{\partial \hat{\rho}_n^{\mathsf{MR}}}{\partial u_{n - p}}$ for any $p \in \mathbb{N}$ always vanish, which is a contradiction. \par
This results shows that in principle, the linear combination of quantum probability readout will yield a function family whose gradient space is much more abundant, because usually the feature number $K-1$ is much larger than the readout qubit number $R$.

\section{Channel equalization: background and training details}
\label{app:ce}

In this Supplementary Note, we provide some more details of the channel equalization task used as an example of time-dependent processing.

\begin{figure}[h]
    \includegraphics[width=1\columnwidth]{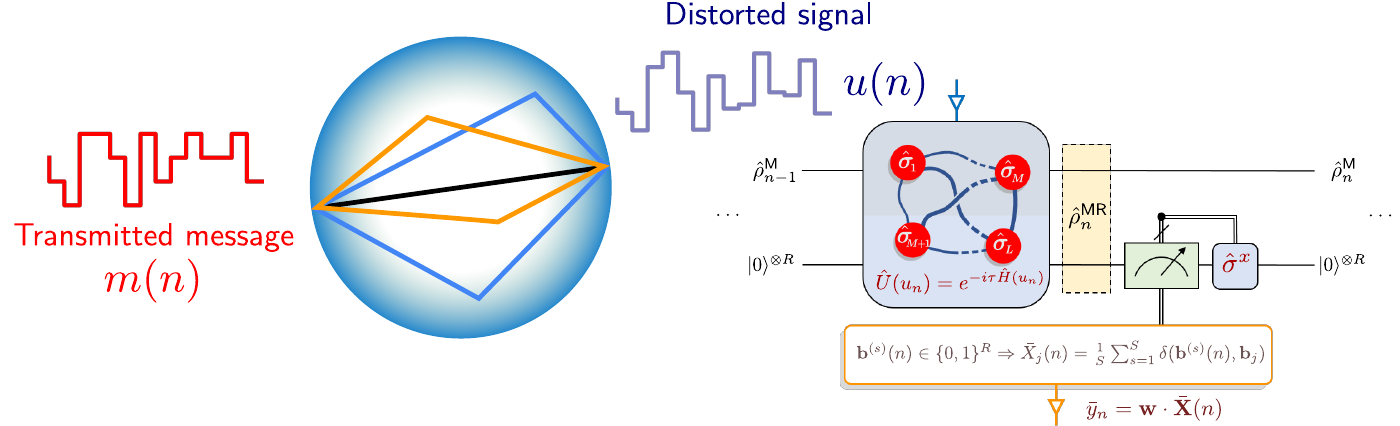}
    \caption{Schematic of the goal of the channel equalization task, and a representative implementation using a QRC under the \nisqrc{} framework. }
    \label{fig:channel_equalization}
\end{figure}

As mentioned in Results' subsection ``Practical machine learning using temporal data'' of the main text, the channel equalization task requires accurately reconstructing a temporally-varying message $m(n)$ from its corrupted copy $u(n)$ after transmission. For the instance we consider, the distortion of the transmitted signal is modeled via the action of a linear kernel $h(n)$, nonlinear mixing $f$ and additive Gaussian noise $\epsilon_0$: 
\begin{align}
    u(n) = f \! \left( \, \sum_{n_1=0}^{7} h(n_1) m(n-n_1) \right) + \epsilon_0. \label{eq:mn2un}
\end{align}
We choose a kernel $h \in \mathbb{R}^8$, whose elements we now specify as $h=[1.0, 0.18, -0.1, 0.091, -0.05, 0.04, 0.03, $ $0.01]$. The nonlinear distortion is modeled by the polynomial $f(x) = x + 0.06x^2 - 0.01x^3$, while the additive noise is parameterized as $\epsilon_0 \in \mathcal{N}\! \left(0, 10^{\frac{-\mathrm{SNR}}{10}} \right)$.
The coefficients in $h$ and $f$ are from the channel equalization task presented in Ref.\,[5]. 
We remove the leading two coefficients of $h$ which represent dependence on future two steps of message symbols. We also increase the nonlinearity in quadratic part of $f$ to make it non-invertible.
Recovering $m(n)$ from $u(n)$ therefore requires a nonzero memory time (to undo the linear kernel), nonlinear processing (to undo the polynomial $f$), and filtering (to remove added noise). In this simulated scenario where $h(n)$ and $f(x)$ are known, the distortion can be inverted up to the additive noise $\epsilon_0$, thus providing a theoretical bound on the minimum achievable error rate. 

We select a $(2+4)$ qubit reservoir, namely with $M=2$ memory qubits and $R=4$ readout qubits. The \nisqrc{} Hamiltonian is as given in Eq.\,(\ref{eq:H0_H1}). We now also detail hyperparameters defining this Hamiltonian for the instance analyzed in Fig.\,\ref{fig:CE_test_error} of the main text. In unit of $1/\tau$, the single-qubit terms are defined by hyperparameters $\eta^z=\varepsilon^z_{\mathrm{rms}} =0.5$ and $\eta^x = \varepsilon^x_{\mathrm{rms}} = 2$. The interaction strengths $J_{i,i'}$ are uniformly sampled from $[0, 1]$, but individual couplings are turned off when analyzing the different QRC connectivities. 

Finally, the $R=4$ readout qubits imply that at each time step $n$ we acquire $K=2^4=16$ readout features $\{\bar{X}_j(n)\}_{j \in [K]}$. A final processing step is the application of a logistic regression layer to these QRC readout features, 
\begin{align}
    y_n = \mathop{\mathrm{argmax}}\limits_{m \in \{-3, -1, 1, 3\}}  \sigma({\mathbf{w}_m \cdot \bar{\mathbf{X}}(n)}).
\end{align}
for computing and minimizing the cross-entropy loss, where $\mathbf{w} \in \mathbb{R}^{4 \times K}$ and $\bar{\mathbf{X}}(n) \in \mathbb{R}^{K}$. The results of testing using this scheme are depicted in Fig.\,\ref{fig:CE_test_error} of the main text. 

We compare \nisqrc{} error rates with two meaningful bounds in Fig.\,\ref{fig:CE_test_error}(a), those derived from the theoretical direct inverse and numerical logistic regression. The lowest possible error rate is that achievable with the direct inverse, which assumes one knows the distorting channel exactly: $y_{\mathrm{DI},n} = \sum_{n_1}h^{-1}(n_1)f^{-1}(u(n-n_1))$, where $h^{-1}$ is the inverse of linear transformation $h$. The noise term $\epsilon_0$ in Supplementary Equation \ref{eq:mn2un} leads to a non-zero error rate for direct inverse.

As an upper bound we consider classical logistic regression applied to the current input value $u(n)$, equivalent to a one-layer perceptron with a softmax activation function.  Specifically, we the output is $y_{\mathrm{LR}, n} = \mathop{\mathrm{argmax}}_m\sigma({\mathbf{w}_{\mathrm{LR},m} \cdot \mathbf{u}(n)})$, where weights $\{\mathbf{w}_{\mathrm{LR},m} \}_{m \in \{\pm1, \pm3\}}$ are trained by minimizing the cross-entropy loss over the same training set used for \nisqrc{}.  Since this results in a linear and memory-less map, improvements in error-rate beyond this upper-bound indicate useful processing done through \nisqrc{}.




\section{IBM Device simulations as a function of qubit coherence times}
\label{app:device_sim}

In this Supplementary Note we provide supplementary simulation results for the IBM device analyzed in the main text. From Fig.\,\ref{fig:CE_expt_error}(b) in the main text, we found that actual device results matched ideal results (in the absence of any losses) very well. Since we are primarily interested in the role of finite qubit coherence times, we now consider the role of a loss model that accounts for finite qubit $T_1$ and $T_2$ times. In particular, we consider normal distributions $T_1 \in \mathcal{N}(\langle T_1 \rangle,\sigma_{T_1})$. and $T_2 \in \mathcal{N}(\langle T_2 \rangle,\sigma_{T_2})$ for the $L=7$ qubit chain. We start with an initial distribution of coherence times consistent with the actual \textit{ibm\_algiers} device from which experimental results are shown in the main text; here $\langle T_1 \rangle \simeq 100~\mu$s, $\langle T_2 \rangle \simeq 170~\mu$s, and $\sigma_{T_1} = \sigma_{T_2} = 10~\mu$s. We then vary the average coherence times across four orders of magnitude (the standard deviations are also scaled by the same factor), and simulate performance of the CE task analyzed in the main text; the resulting error rates are plotted in Supplementary Figure \ref{fig:device_sim}.

We note that for coherence times that are an order of magnitude shorter than the typical device coherence times, the CE task performance is essentially unaffected. In fact, even for very low coherence times of $\langle T_1 \rangle \simeq 1~\mu$s, around two orders of magnitude shorter than device lifetimes, a nontrivial I/O map is retained by the \nisqrc{} algorithm and the considered instance of the CE task can still be performed (albeit with a testing error rate that now is marginally worse than that of single-step logistic regression). For even lower coherence times the I/O map will ultimately become trivial as errors uncorrelated with the input encoding start to overwhelm the dynamics of the system, and hence any outputs extracted from it.


\begin{figure}[t]
    \centering
    \includegraphics[width = 0.6\linewidth]{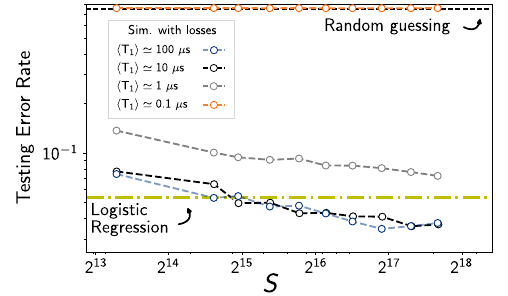}
    \caption{Testing error rates for the CE task in the main text, Fig.\,\ref{fig:CE_expt_error} as a function of number of shots $\NS$ using simulations of the \textit{ibm\_algiers}, now for varying qubit coherence times. Details on coherence time values are provided in the text. For comparison, we plot the testing error rate of logistic regression (yellow line), as well as random guessing (black dashed line).}
    \label{fig:device_sim}
\end{figure}



\section{IBM Device experiments under controlled delays}
\label{app:device_delay}

Fig.\,\ref{fig:CE_expt_error} of the main test shows the results of performing the CE task on the IBMQ device \textit{ibm\_algiers}.  With the circuit we have employed, the total run time $T_{\rm run}$ approaches the average qubit $T_1$ times on this device. In principle, the \nisqrc{} algorithm enables $T_{\rm run}$ to exceed $T_1$ indefinitely provided $T_1 > n_{\rm M}^0$ required for the specific instance of the CE task. However, due to limitations on the classical processing backend, the experiments are unable to be run for longer messages than $N=20$ as of present, so that $T_{\rm run}$ cannot be increased naturally by increasing $N$.

In this Supplementary Note, we present the results of an experiment used to artificially lengthen the total circuit run time $T_{\rm run}$ by introducing controlled delays to the circuit. The circuit schematic we implement is shown in Supplementary Figure \ref{fig:device_delay}(a), with the grey block indicating delays added after each set of gate applications, measurement, and reset operations, except after the final measurement. We consider delays that are typically much larger than the total time $\tau$ in each unit of evolution under \nisqrc{}. We emphasize that during the delay time, the qubits forming the QRC can experience decay due to their finite lifetime. In the absence of delays, the circuit run time is $T_{\rm run} = 117~\mu$s, as indicated in the main text. By introduce a delay of $T_{\rm delay} = 20~\mu$s or $T_{\rm delay} = 40~\mu$s per unit (significantly longer than the unit evolution time $\tau$), the run time can be extended to $T_{\rm run} = 497~\mu$s or $T_{\rm run} = 877~\mu$s respectively; the latter is almost an order of magnitude larger than the mean $T_1 = 155~\mu$s.

The testing error rate achieved is shown in Supplementary Figure \ref{fig:device_delay}(b). Here we show the performance cumulatively averaged over $P$ permutations of the training and testing datasets, a standard cross-validation technique to remove fluctuations in performance when having access to only small datasets, and one we use for all results in the main text. We note that even with significantly longer run times, the device is able to beat logistic regression at the CE task. Increasing the delay from $20~\mu$s to $40~\mu$s per unit also does not significantly effect the performance, further highlighting the ability of \nisqrc{} to overcome $T_1$ limitations on run time.

We note that the absolute performance shown in Supplementary Figure \ref{fig:device_delay}(b) is achieved with a larger $\NS \simeq 2^{16}$ than the largest value shown in Fig.\,\ref{fig:CE_expt_error} of the main text. The results in this Supplementary Note are calculated using data obtained several months after the data in Fig.\,\ref{fig:CE_expt_error} of the main text. The slight reduction in performance observed can be attributed to drift in the device over this time frame.


\begin{figure}[t]
    \centering
    \includegraphics[width = 1\linewidth]{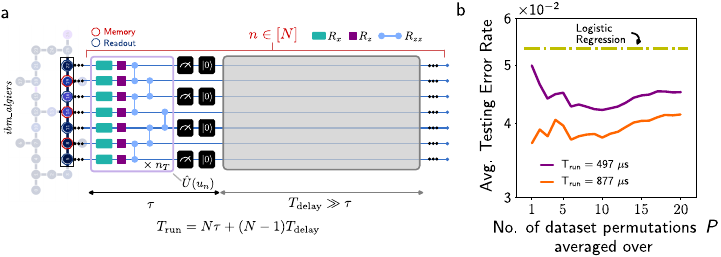}
    \caption{\textbf{(a)} Experimental testing error rates for the CE task in the main text, Fig.\,\ref{fig:CE_expt_error} under the inclusion of controlled delays to increase total circuit run time much beyond individual qubit $T_1$ times, $T_{\rm run} \gg T_1$. Experiments are once again run on \textit{ibm\_algiers}, with mean qubit $T_1$ times of $\langle T_1 \rangle = 155~\mu $s. \textbf{(b)} For comparison, we plot the testing error rate of logistic regression (yellow line).}
    \label{fig:device_delay}
\end{figure}


\section{Lists of device parameters}
\label{app:device_parameter}

Here we list device parameters used for producing Fig.\,\ref{fig:CE_expt_error}(b) in the main text. All $R_z$ gates are implemented as error-free virtual rotation $z$ gates. Since all qubits used form a line, all the CNOT gate errors in the table are indexed by the qubits with small numbering in each control-target pair, therefore CNOT error is not applicable to the qubit 22. The averaged $T_1, T_2$ time over the three experiments are $T_1 \approx 124~\mu$s and $T_2 \approx 91~\mu$s. 
 
\begin{table}[ht]
    \centering
    \begin{tabularx}{1.0\textwidth}{ >{\centering\arraybackslash}X | *{7}{>{\centering\arraybackslash}X}}
    \hline
    Qubit & $5$ & $8$ & $11$ & $14$ & $16$ & $19$ & $22$ \\ \hline
    $T_1$ ($\mu$s) & $159 \pm 43$ & $142 \pm 29$ & $144 \pm 32$ & $164 \pm 20$ & $160 \pm 35$ & $127 \pm 38$ & $147 \pm 42$ \\ \hline
    $T_2$ ($\mu$s) & $96 \pm 34$ & $231 \pm 30$ & $32 \pm 7$ & $97 \pm 4$ & $68 \pm 1$ & $29 \pm 7$ & $163 \pm 28$ \\ \hline
    $\sqrt{X}$ error (\%) & $0.026 \pm 0.006$ & $0.043 \pm 0.027$ & $0.081 \pm 0.060$ & $0.037 \pm 0.018$ & $0.037 \pm 0.005$ & $0.050 \pm 0.016$ & $0.020 \pm 0.005$ \\ \hline
    \makecell{CNOT error \\ to next (\%)} & $0.936 \pm 0.571$ & $1.028 \pm 0.569$ & $0.762 \pm 0.228$ & $6.850 \pm 0.980$ & not reported & $1.162 \pm 0.251$ &  N/A \\ \hline
    \makecell{Readout \\ error (\%)} & $0.746 \pm 0.150$ & $0.855 \pm 0.093$ & $1.412 \pm 0.617$ & $1.865 \pm 0.528$ & $26.292 \pm 19.676$ & $9.675 \pm 0.789$ & $1.129 \pm 0.274$ \\ \hline
    \makecell{Readout \\ length ($\mu$s)} & $0.857 \backslash 0.910$ & $0.857 \backslash 0.910$ & $0.857 \backslash 0.910$ & $0.857 \backslash 0.910$ & $0.857 \backslash 0.910$ & $0.857 \backslash 0.910$ & $0.857 \backslash 0.910$ \\ \hline
    \end{tabularx}
    \caption{Device parameters for connected QRC with mid-circuit measurement and deterministic reset (purple line in Fig.\,\ref{fig:CE_expt_error}(b) of the main text). The calibrations to the CNOT gates between qubit 16 and qubit 19 are not successfully fitted, hence not reported by \textit{ibqm\_algiers} device. The re-calibration of readout length on July 14th, 2023, caused the pre$\backslash$post values $0.857 \backslash 0.910~\mu$s, therefore the experiments for different shots $S$ have different readout lengths.}
    \label{tab:my_label_02}
\end{table}

\begin{table}[t]
    \centering
    \begin{tabularx}{1.0\textwidth}{ >{\centering\arraybackslash}X | *{7}{>{\centering\arraybackslash}X}}
    \hline
    Qubit & $5$ & $8$ & $11$ & $14$ & $16$ & $19$ & $22$ \\ \hline
    $T_1$ ($\mu$s) & $63 \pm 22$ & $125 \pm 29$ & $115 \pm 25$ & $139 \pm 32$ & $90 \pm 17$ & $102 \pm 19$ & $120 \pm 24$ \\ \hline
    $T_2$ ($\mu$s) & $93 \pm 12$ & $192 \pm 52$ & $26 \pm 2$ & $66 \pm 16$ & $9 \pm 1$ & $51 \pm 3$ & $146 \pm 32$ \\ \hline
    $\sqrt{X}$ error (\%) & $0.028 \pm 0.005$ & $0.020 \pm 0.002$ & $0.067 \pm 0.025$ & $0.028 \pm 0.013$ & $0.070 \pm 0.010$ & $0.030 \pm 0.006$ & $0.017 \pm 0.001$ \\ \hline
    \makecell{CNOT error \\ to next (\%)} & $0.591 \pm 0.074$ & $2.255 \pm 1.018$ & $2.256 \pm 1.139$ & $1.655 \pm 0.254$ & $3.386 \pm 0.491$ & $0.848 \pm 0.058$ &  N/A \\ \hline
    \makecell{Readout \\ error (\%)} & $0.803 \pm 0.108$ & $0.771 \pm 0.051$ & $1.314 \pm 0.349$ & $3.561 \pm 0.282$ & $6.053 \pm 1.643$ & $2.546 \pm 0.097$ & $0.710 \pm 0.084$ \\ \hline
    \makecell{Readout \\ length ($\mu$s)} & $0.910$ & $0.910$ & $0.910$ & $0.910$ & $0.910$ & $0.910$ & $0.910$ \\ \hline
    \end{tabularx}
    \caption{Device parameters for split QRC with mid-circuit measurement and deterministic reset (brown line in Fig.\,\ref{fig:CE_expt_error}(b) of the main text). }
    \label{tab:my_label_03}
\end{table}

\clearpage

\begin{table}[t]
    \centering
    \begin{tabularx}{1.0\textwidth}{ >{\centering\arraybackslash}X | *{7}{>{\centering\arraybackslash}X}}
    \hline
    Qubit & $5$ & $8$ & $11$ & $14$ & $16$ & $19$ & $22$ \\ \hline
    $T_1$ ($\mu$s) & $63 \pm 29$ & $137 \pm 15$ & $118 \pm 20$ & $154 \pm 26$ & $114 \pm 13$ & $85 \pm 25$ & $133 \pm 13$ \\ \hline
    $T_2$ ($\mu$s) & $75 \pm 21$ & $205 \pm 39$ & $26 \pm 2$ & $76 \pm 8$ & $10 \pm 1$ & $47 \pm 7$ & $159 \pm 30$ \\ \hline
    $\sqrt{X}$ error (\%) & $0.035 \pm 0.010$ & $0.018 \pm 0.001$ & $0.072 \pm 0.054$ & $0.017 \pm 0.002$ & $0.075 \pm 0.019$ & $0.030 \pm 0.002$ & $0.017 \pm 0.002$ \\ \hline
    \makecell{CNOT error \\ to next (\%)} & $0.742 \pm 0.321$ & $0.954 \pm 0.320$ & $1.020 \pm 0.264$ & $1.564 \pm 0.156$ & $3.532 \pm 0.605$ & $0.840 \pm 0.058$ &  N/A \\ \hline
    \makecell{Readout \\ error (\%)} & $1.014 \pm 0.464$ & $0.770 \pm 0.088$ & $1.299 \pm 0.183$ & $3.438 \pm 0.418$ & $5.556 \pm 1.982$ & $2.473 \pm 0.133$ & $0.818 \pm 0.142$ \\ \hline
    \makecell{Readout \\ length ($\mu$s)} & $0.910$ & $0.910$ & $0.910$ & $0.910$ & $0.910$ & $0.910$ & $0.910$ \\ \hline
    \end{tabularx}
    \caption{Device parameters for connected QRC with mid-circuit measurement, but without deterministic reset (green line in Fig.\,\ref{fig:CE_expt_error}(b) of the main text). }
    \label{tab:my_label_05}
\end{table}


\section*{Supplementary References}

\begin{itemize}
    \item[{[1]}] F. Hu, G. Angelatos, S. A. Khan, M. Vives, E. Türeci, L. Bello, G. E. Rowlands, G. J. Ribeill, and H. E. Türeci, Tackling sampling noise in physical systems for machine learning applications: Fundamental limits and eigentasks, Physical Review X \textbf{13}, 041020 (2023).
    \item[{[2]}] J. Chen, H. I. Nurdin, and N. Yamamoto, Temporal Information Processing on Noisy Quantum Computers, Physical Review Applied \textbf{14}, 024065 (2020).
    \item[{[3]}] B. Skinner, J. Ruhman, and A. Nahum, Measurement-induced phase transitions in the dynamics of entanglement, Physical Review X \textbf{9}, 031009 (2019).
    \item[{[4]}] T. Yasuda, Y. Suzuki, T. Kubota, K. Nakajima, Q. Gao, W. Zhang, S. Shimono, H. I. Nurdin, and N. Yamamoto, Quantum reservoir computing with repeated measurements on superconducting devices, arXiv:2310.06706 [quant-ph] (2023).
    \item[{[5]}] H. Jaeger and H. Haas, Harnessing Nonlinearity: Predicting Chaotic Systems and Saving Energy in Wireless Communication, Science \textbf{304}, 78 (2004).
\end{itemize}
\end{widetext}

\end{document}